\documentclass[aps,prd,preprintnumbers,showpacs,superscriptaddress,nofootinbib,amsmath,amssymb,floats,floatfix,showkeys,notitlepage,longbibliography]{revtex4-1}
\usepackage{hyphenat}
\usepackage[table]{xcolor} 
\usepackage{array} 
\renewcommand{\arraystretch}{1.5} 
\definecolor{lightblue}{rgb}{0.62, 0.83, 0.96} 
\definecolor{lightgray}{rgb}{0.93, 0.97, 0.99} 
\usepackage{orcidlink}
\usepackage{comment}
\usepackage{graphicx}
\usepackage{subfigure}
\usepackage{palatino}
\usepackage{float}
\usepackage[commandnameprefix=always]{changes}
\usepackage{hyperref}
\hypersetup{colorlinks=true,linkcolor=red,urlcolor=red,citecolor=red}
\usepackage[toc,page]{appendix}
\usepackage[normalem]{ulem}
\usepackage{lipsum}
\usepackage{graphicx}
\usepackage{subfigure}
\usepackage{palatino}
\usepackage{sans}
\usepackage{adjustbox}
\usepackage{latexsym}
\usepackage{amsmath}
\usepackage{amssymb}
\usepackage{amsfonts}
\usepackage{dcolumn}
\usepackage{bm}
\usepackage{tikz}
\usepackage{bigints}
\usepackage{array,tabularx,multirow,booktabs}
\usepackage[tracking=true]{microtype}
\SetTracking{}{500}
\SetTracking{encoding={*}, shape=sc}{40}
\allowdisplaybreaks
\usepackage{adjustbox}
\usepackage{latexsym}
\usepackage{amsmath}
\usepackage{amssymb}
\usepackage{amsfonts}
\usepackage{dcolumn}
\usepackage{bm}
\usepackage{tikz}
\usepackage{bigints}
\usepackage{array,tabularx,multirow,booktabs}
\usepackage[tracking=true]{microtype}
\usepackage{color}
\usepackage{todonotes}
\allowdisplaybreaks
\begin{document}

\title{Beyond Extremality: Weak Gravity Conjecture Constraints on Gravitational Lensing in Gravity’s Rainbow}

\author{Saeed Noori Gashti
\orcidlink{0000-0001-7844-2640}}
\email{sn.gashti@du.ac.ir}
\affiliation{School of Physics, Damghan University, Damghan 3671645667, Iran}

\author{Behnam Pourhassan \orcidlink{0000-0003-1338-7083}
}
\email{b.pourhassan@du.ac.ir}
\affiliation{School of Physics, Damghan University, Damghan 3671645667, Iran}
\affiliation{Center for Theoretical Physics, Khazar University, 41 Mehseti Street, Baku, AZ1096, Azerbaijan}

\author{\.{I}zzet Sakall{\i}
\orcidlink{0000-0001-7827-9476}
}
\email{izzet.sakalli@emu.edu.tr}
\affiliation{Physics Department, Eastern Mediterranean
University, Famagusta, 99628 North Cyprus, via Mersin 10, T\"{u}rkiye}

\begin{abstract}
We investigate the constraints imposed by the Weak Gravity Conjecture (WGC) on gravitational lensing in gravity’s rainbow, focusing in particular on scenarios beyond extremality and on the interplay between the WGC and the Weak Cosmic Censorship Conjecture (WCCC) in the context of Reissner–Nordström–Anti-de Sitter black holes modified by rainbow gravity. Using topological methods, we first analyze the configuration of photon spheres and confirm that unstable circular photon spheres with topological charge $(\omega = -1)$ exist outside the event horizon throughout the parameter space, thereby verifying the simultaneous validity of both the WGC and the WCCC. The rainbow functions $f(\varepsilon)$ and $g(\varepsilon)$, which encode Planck-scale corrections through the energy ratio $(\varepsilon = E/E_P)$, modify both the spacetime metric and the extremality bound. We derive the corresponding modified extremal charge-to-mass ratio, $(q^2/m^2) > (Q^2/M^2)_{\text{ext}}$, and show that gravity’s rainbow offers a natural mechanism for reconciling these two fundamental conjectures. By applying the Gauss–Bonnet theorem in conjunction with Jacobi–Maupertuis optical geometry, we compute the weak deflection angles for both photons and massive particles to second order. The rainbow function $g(\varepsilon)$ appears with powers $(g^{-2})$ and $(g^{-4})$, enhancing the deflection angle when $g(\varepsilon) < 1$, while $f(\varepsilon)$ influences only the charge-dependent contributions. At extremality, the deflection angle becomes independent of $f(\varepsilon)$, yielding a universal prediction that can be tested without specifying the form of the rainbow functions. We further find that super-extremal configurations exhibit stronger lensing effects than extremal black holes, suggesting a potential observational discriminator between WGC-satisfying naked singularities and WCCC-preserving black holes. These findings position gravitational lensing as a promising tool for probing quantum-gravity phenomenology with upcoming observational facilities.
\end{abstract}

\date{\today}



\maketitle
\tableofcontents
\newpage
\section{Introduction}\label{isec1}

The quest for a consistent theory of quantum gravity remains one of the most profound challenges in modern theoretical physics, drawing sustained attention from researchers across high-energy physics, cosmology, and mathematical physics \cite{01,02}. At its core, quantum gravity seeks to reconcile the principles of quantum mechanics with those of general relativity (GR), thereby providing a unified description of spacetime at the smallest length scales where both gravitational and quantum effects become significant. While a complete formulation of quantum gravity remains elusive, substantial progress has been made through various approaches, including string theory, loop quantum gravity, asymptotic safety, and causal set theory \cite{03,04,05}. Among these diverse programs, the swampland initiative has emerged as a particularly influential framework for constraining effective field theories that could arise from ultraviolet (UV) complete quantum gravity theories \cite{1,2,3,4,5,6,7,8}.

The swampland program rests on a fundamental observation: not all effective field theories that appear mathematically self-consistent at low energies can be embedded into a UV-complete theory of quantum gravity, such as string theory \cite{1,2}. This realization leads to a natural classification of theories into two categories. The landscape comprises theories that admit such an embedding and can thus be considered viable candidates for low-energy limits of quantum gravity. In contrast, the swampland contains theories that, despite their apparent internal consistency, fail to arise from any UV-complete gravitational theory and must therefore be excluded from physical consideration. The swampland conjectures provide a set of proposed criteria---grounded in black hole (BH) thermodynamics, holography, and string theory constructions---that aim to distinguish between these two classes \cite{1,2,3,4,5,6,7,8}.

The physical intuition underlying these conjectures draws from multiple foundational areas. BH thermodynamics, particularly the Bekenstein-Hawking entropy and information paradox, offers crucial guidance regarding the quantum nature of gravitational degrees of freedom. The holographic principle, most precisely realized through the Anti-de Sitter/Conformal Field Theory (AdS/CFT) correspondence, provides a non-perturbative definition of quantum gravity in certain spacetime backgrounds and establishes deep connections between gravitational dynamics and boundary quantum field theories \cite{1,2,3,4,5,6,7,8}. String theory dualities and explicit compactification constructions supply concrete examples that ground the swampland criteria in calculable frameworks. Together, these threads weave a rich tapestry of constraints that any consistent theory of quantum gravity must satisfy, with implications spanning early universe cosmology, dark energy phenomenology, and particle physics beyond the Standard Model \cite{a,b,f,g,h,i,j,k,l,m,n,o,p,q,r,s,t,u,v,w,x,y,z,aa,bb,cc,dd,ee,ff,gg,hh,ii,jj,kk,ll,mm,nn,oo,pp,qq,rr,ss,tt,uu,vv,ww,xx,yy,zz,aaa,bbb,ccc,ddd,eee,fff,ggg,hhh,iii,jjj,kkk,lll,rrr,sss,ttt,uuu,vvv,www,xxx,zzz,zzzz}.

A cornerstone of the swampland program is the WGC, which asserts that in any consistent theory of quantum gravity coupled to a $U(1)$ gauge field, there must exist particles whose charge-to-mass ratio satisfies $q/m \geq 1$ in appropriate units. Equivalently, for charged BHs, the WGC demands the existence of states that can facilitate BH decay, preventing the formation of stable extremal remnants that would conflict with expectations from BH thermodynamics and holography \cite{9,10}. The conjecture effectively encodes the principle that ``gravity is the weakest force''---electromagnetic repulsion must overcome gravitational attraction for sufficiently charged particles. The WGC has found applications extending well beyond its original formulation, influencing discussions of axion physics, cosmic inflation, dark matter candidates, and even the cosmological constant problem \cite{9,10}.

Parallel to these developments in quantum gravity, the WCCC, originally proposed by Penrose, addresses a complementary question in classical gravitational physics . The WCCC posits that singularities produced by gravitational collapse are generically hidden behind event horizons, rendering them invisible to distant observers and thereby preserving the predictability of spacetime evolution. Without such protection, naked singularities would expose regions of arbitrarily high curvature to external observation, leading to breakdowns in determinism and potentially invalidating the entire framework of classical GR as a predictive theory. The WCCC has been tested extensively through gedanken experiments involving the capture of test particles by BHs, accretion of charged matter, and various perturbation analyses, with results generally supporting its validity under physically reasonable conditions \cite{10000,11,12}.

However, a fundamental tension emerges when the WGC and WCCC are considered together in the context of charged BHs, particularly the Reissner--Nordstr\"{o}m (RN) solution describing static, spherically symmetric BHs carrying electric charge \cite{jjj,kkk}. The RN metric possesses an event horizon only when the charge-to-mass ratio satisfies $Q/M \leq 1$; beyond this threshold, the horizon structure degenerates and a naked singularity appears, violating the WCCC. Conversely, the WGC requires the existence of particles with $q/m > 1$, which for standard RN BHs implies $q/m > (Q/M)_{\rm ext}$. If such particles exist and can be absorbed by a near-extremal BH, the resulting configuration could become super-extremal ($Q > M$), exposing a naked singularity and contradicting the WCCC. This apparent conflict between two fundamental conjectures---one rooted in quantum gravity considerations and the other in classical gravitational dynamics---motivates a deeper investigation into their mutual compatibility and the conditions under which both can be simultaneously satisfied \cite{45'm,45mmm,45m}.

Recent theoretical work has identified several mechanisms that could reconcile the WGC and WCCC. The inclusion of additional matter fields, such as quintessence or dark matter components, modifies the extremality bound and can prevent overcharging \cite{jjj,kkk,45'm,45mmm,45m}. Similarly, spacetimes with cosmological constants (either positive or negative) alter the horizon structure and shift the critical charge-to-mass ratio at which extremality occurs. Modified gravity theories, including those with higher-curvature corrections, introduce additional parameters that influence both the WGC bound and the conditions for cosmic censorship. These investigations suggest that the tension between the WGC and WCCC may be an artifact of oversimplified models, and that more realistic scenarios incorporating quantum gravity effects naturally accommodate both conjectures.

Gravity's rainbow represents one such framework incorporating quantum gravity modifications into the spacetime geometry \cite{033,034}. Originally motivated by doubly special relativity and deformed dispersion relations arising in loop quantum gravity and string theory contexts, gravity's rainbow introduces energy-dependent rainbow functions $f(\varepsilon)$ and $g(\varepsilon)$ that modify the metric components according to the probe particle's energy relative to the Planck scale \cite{035,036}. These functions satisfy the infrared limit $\lim_{\varepsilon \to 0} f(\varepsilon) = \lim_{\varepsilon \to 0} g(\varepsilon) = 1$, recovering standard GR when probe energies are much smaller than the Planck energy $E_P$. The rainbow framework has been applied to numerous BH solutions, revealing modifications to thermodynamic properties, quasi-normal modes, Hawking radiation spectra, and geodesic structures \cite{037,038}. Crucially for the present work, gravity's rainbow alters the extremality condition for charged BHs, providing a natural mechanism for shifting the WGC bound while preserving cosmic censorship.

Gravitational lensing serves as one of the most powerful observational probes for testing gravitational theories and constraining BH parameters \cite{039,040}. The deflection of light by massive objects, first predicted by GR and confirmed during the 1919 solar eclipse, has evolved into a precision tool encompassing weak lensing surveys, strong lensing time delays, and BH shadow observations \cite{041}. The Gauss-Bonnet (GB) theorem, combined with the Jacobi-Maupertuis optical geometry formalism pioneered by Gibbons and Werner, provides an elegant geometric framework for calculating deflection angles by integrating Gaussian curvature over appropriately defined optical manifolds \cite{042,043}. This approach has been extended to numerous BH spacetimes, including those with rotation, charge, and various matter surroundings, enabling analytical treatment of lensing phenomena across diverse theoretical scenarios \cite{044,045}.

Photon spheres (PSs)---the unstable circular orbits of massless particles around compact objects---provide additional geometric information about BH spacetimes and connect naturally to both shadow observations and gravitational lensing. The topological classification of PSs, assigning winding numbers or topological charges that encode stability properties, has emerged as a fruitful approach for characterizing BH geometries. The existence of an unstable PS outside the event horizon serves as a necessary condition for BH nature, as naked singularities typically lack such structures or exhibit qualitatively different PS configurations. This connection makes PS analysis a valuable complement to direct horizon studies when assessing whether a given spacetime represents a BH or a naked singularity \cite{45'm,45mmm,45m,46m,47m,48m,49m,50m}.

The primary motivation for the present work stems from the desire to establish observational signatures that can discriminate between WGC-compatible and WCCC-compatible configurations in gravity's rainbow. While both conjectures address fundamental questions about quantum gravity and spacetime structure, they have remained largely theoretical constructs with limited connection to observable phenomena. Gravitational lensing, with its sensitivity to spacetime geometry and upcoming observational capabilities through facilities such as the Square Kilometre Array (SKA) and next-generation Event Horizon Telescope (ngEHT), offers a promising pathway toward empirical tests. By deriving explicit deflection angle formulas for Reissner--Nordstr\"{o}m--AdS (R-N-AdS) BHs modified by gravity's rainbow, we aim to identify parameter regimes where WGC and WCCC predictions diverge and to establish concrete observational targets for future campaigns.

Our objectives in this study are fourfold. First, we construct the R-N-AdS BH solution within the gravity's rainbow framework, incorporating power-law nonlinear electrodynamics characterized by the nonlinearity parameter $p$. Second, we analyze the PS structure using topological methods to verify that BH configurations persist across the rainbow parameter space and that the WGC bound is satisfied simultaneously with PS existence. Third, we derive the weak deflection angle for both photons and massive particles using the GB theorem approach, obtaining explicit expressions in terms of the rainbow functions $f(\varepsilon)$ and $g(\varepsilon)$. Fourth, we investigate the behavior of lensing observables as the BH transitions from sub-extremal to extremal and super-extremal configurations, revealing how the competition between WGC and WCCC manifests in observable deflection angle signatures.

The paper is organized as follows. In Section~\ref{isec2}, we present the R-N-AdS BH solution modified by gravity's rainbow, deriving the metric function and analyzing the PS structure through topological methods. We demonstrate the simultaneous satisfaction of WGC bounds and PS existence across the parameter space. Section~\ref{isec3} develops the gravitational lensing formalism based on the GB theorem and Jacobi-Maupertuis optical geometry. We derive the Gaussian curvature of the optical manifold, compute weak deflection angles to second order, and examine the implications for WGC constraints at extremality. We also present the universal $f$-independence of extremal deflection angles and contrast lensing signatures across the WGC/WCCC boundary. Finally, Section~\ref{isec4} summarizes our findings, discusses observational prospects, and outlines directions for future research.

\section{R-N-AdS BH Solution in Gravity's Rainbow Framework} \label{isec2}

The interplay between nonlinear electrodynamics (NLED) and modified dispersion relations offers a fertile ground for exploring quantum gravity effects on charged BH spacetimes. Gravity's rainbow, originally motivated by doubly special relativity and loop quantum gravity considerations, introduces energy-dependent modifications to the spacetime metric that become significant near the Planck scale. In this section, we construct the R-N-AdS BH solution within this framework and examine its structural properties.

\subsection{Action and Field Equations}

We begin with the gravitational action describing four-dimensional BHs coupled to NLED within Einstein gravity \cite{m1,m2}:
\begin{equation}\label{m1}
I = -\frac{1}{16\pi} \int d^4 x \sqrt{-g} \left(R - 2\Lambda + \mathcal{L}(F)\right),
\end{equation}
where $R$ is the Ricci scalar curvature, and $\Lambda$ denotes the cosmological constant. For AdS spacetimes, we have $\Lambda = -3/l^2$, with $l$ being the AdS radius. The electromagnetic Lagrangian density $\mathcal{L}(F)$ depends on the Maxwell invariant $F = F_{\mu\nu}F^{\mu\nu}$, where the field strength tensor is defined as $F_{\mu\nu} = \partial_\mu A_\nu - \partial_\nu A_\mu$ with $A_\mu$ representing the gauge potential. Following the power-law NLED formulation \cite{m2,m3}, we adopt:
\begin{equation}\label{m2}
\mathcal{L}(F) = (-F)^p,
\end{equation}
where the exponent $p$ serves as the nonlinearity parameter that controls deviations from standard Maxwell theory.

Varying the action \eqref{m1} with respect to the metric tensor $g_{\mu\nu}$ produces Einstein's field equations:
\begin{equation}\label{m3}
R_{\mu\nu} - \frac{1}{2} R g_{\mu\nu} + \Lambda g_{\mu\nu} = \frac{1}{2} g_{\mu\nu} (-F)^p + 2p (-F)^{p-1} F_{\mu\alpha} F_{\nu}^{\alpha}.
\end{equation}
The variation with respect to the electromagnetic potential $A_\mu$ yields the generalized Maxwell equations:
\begin{equation}\label{m4}
\partial_\mu \left(\sqrt{-g} \mathcal{L}'(F) F^{\mu\nu}\right) = 0,
\end{equation}
with $\mathcal{L}_F = d\mathcal{L}/dF$. For the configuration under consideration, only the radial-temporal component $F_{tr}$ of the electromagnetic field tensor is nonvanishing.

\subsection{Rainbow-Modified Metric Structure}

To incorporate gravity's rainbow effects, we consider a static, spherically symmetric, and energy-dependent spacetime characterized by the line element \cite{m4}:
\begin{equation}\label{m5}
ds^2 = -\frac{V(r)}{f^2(\varepsilon)} dt^2 + \frac{1}{g^2(\varepsilon)} \left(\frac{dr^2}{V(r)} + r^2 d\Omega^2 \right),
\end{equation}
where $d\Omega^2 = d\theta^2 + \sin^2 \theta\, d\phi^2$ represents the standard two-sphere metric. The rainbow functions $f(\varepsilon)$ and $g(\varepsilon)$ modify the temporal and spatial sectors of the metric, respectively. Here, the dimensionless parameter $\varepsilon = E/E_P$ quantifies the ratio of a test particle's energy $E$ to the Planck energy $E_P$. These rainbow functions must satisfy the infrared limit:
\begin{equation}\label{m6}
\lim_{\varepsilon \to 0} f(\varepsilon) = 1, \quad \lim_{\varepsilon \to 0} g(\varepsilon) = 1,
\end{equation}
which ensures that standard GR is recovered when probe energies are much smaller than the Planck scale. Various phenomenologically motivated forms for these functions have been proposed in the literature \cite{m5,m6}:
\begin{align}
f(\varepsilon) &= 1, \quad g(\varepsilon) = \sqrt{1 - \eta \varepsilon^n}, \label{rainbow1}\\
f(\varepsilon) &= \frac{e^{\xi \varepsilon} - 1}{\xi \varepsilon}, \quad g(\varepsilon) = 1, \label{rainbow2}\\
f(\varepsilon) &= g(\varepsilon) = \frac{1}{1 - \lambda \varepsilon}, \label{rainbow3}
\end{align}
where $\eta$, $n$, $\xi$, and $\lambda$ are dimensionless constants that parameterize the specific rainbow model. Each choice leads to distinct physical predictions for high-energy phenomena near the BH.

\subsection{Electromagnetic Field Configuration}

Assuming the gauge potential possesses only a temporal component $A_t(r)$, the nonzero electromagnetic field strength reads:
\begin{equation}\label{m7}
F_{tr} = -\partial_r A_t(r),
\end{equation}
and the Maxwell invariant takes the form:
\begin{equation}\label{m8}
F = -2 f^2(\varepsilon) g^2(\varepsilon) (F_{tr})^2 = -2 f^2(\varepsilon) g^2(\varepsilon) (A_t'(r))^2.
\end{equation}
Substituting this expression into Eq.~\eqref{m4} leads to a nonlinear differential equation governing $A_t(r)$:
\begin{equation}\label{m9}
r (A_t'(r))^{2p - 2} \left[ (2p - 1) r A_t''(r) + 2A_t'(r) \right] = 0, \quad p \neq \frac{1}{2}.
\end{equation}
The general solution separates into two branches depending on the value of $p$:
\begin{equation}\label{m10}
A_t(r) =
\begin{cases}
-q \ln\left(\dfrac{r}{l}\right), & p = \dfrac{3}{2}, \\[10pt]
-q\left(\dfrac{2p - 1}{2p - 3}\right) r^{\frac{2p - 3}{2p - 1}}, & \dfrac{1}{2} < p < \dfrac{3}{2},
\end{cases}
\end{equation}
where $q$ is an integration constant associated with the BH's electric charge. The corresponding electromagnetic tensor component becomes:
\begin{equation}\label{m11}
F_{tr} =
\begin{cases}
\dfrac{q}{r}, & p = \dfrac{3}{2}, \\[8pt]
q\, r^{-\frac{2}{2p - 1}}, & \dfrac{1}{2} < p < \dfrac{3}{2}.
\end{cases}
\end{equation}
When $p=1$, the standard R-N-AdS BH solution is recovered, corresponding to linear Maxwell electrodynamics. The constraint $1/2 < p \leq 3/2$ guarantees that $A_t(r)$ remains finite at spatial infinity, preserving the physical viability of the solution.

\subsection{Derivation of the Metric Function}

Substituting the metric ansatz \eqref{m5} and the electromagnetic solutions into the gravitational field equations yields the radial components:
\begin{equation}
\begin{cases}
e_{tt} = e_{rr} = g^{2}(\varepsilon) \left( V''(r) + \dfrac{2 V'(r)}{r} \right) + 2\Lambda - 2 \sqrt{2} q^{3}_\epsilon \, r^{-3} = 0, & p = \dfrac{3}{2}, \\[12pt]
e_{tt} = e_{rr} = g^{2}(\varepsilon) \left( V''(r) + \dfrac{2 V'(r)}{r} \right) + 2\Lambda - 2^p q^{2p}_\epsilon \, r^{-\frac{4p}{2p-1}} = 0, & \dfrac{1}{2} < p < \dfrac{3}{2},
\end{cases}
\label{eq:ett_err}
\end{equation}
together with the angular components:
\begin{equation}
\begin{cases}
e_{\theta \theta} = e_{\varphi \varphi} = 2 g^{2}(\varepsilon) \left( \dfrac{V'(r)}{r} + \dfrac{V(r) - 1}{r^{2}} \right) + 2\Lambda + 4\sqrt{2} q^{3}_ \epsilon \, r^{-3} = 0, & p = \dfrac{3}{2}, \\[12pt]
e_{\theta \theta} = e_{\varphi \varphi} = 2 g^{2}(\varepsilon) \left( \dfrac{V'(r)}{r} + \dfrac{V(r) - 1}{r^{2}} \right) + 2\Lambda + (2p - 1) 2^p q^{2p}_\epsilon \, r^{-\frac{4p}{2p-1}} = 0, & \dfrac{1}{2} < p < \dfrac{3}{2}.
\end{cases}
\label{eq:etheta_ephi}
\end{equation}
Here, $e_{\mu\nu}$ represents the components of the Einstein tensor minus the stress-energy contributions, and we use the shorthand notation $q_\epsilon \equiv q \, f(\varepsilon) g(\varepsilon)$. These equations are not independent; one can verify that:
\begin{equation}\label{m12}
e_{tt} = \left(1 + \frac{r}{2} \frac{d}{dr}\right) e_{\theta \theta},
\quad \text{for} \quad \frac{1}{2} < p \leq \frac{3}{2}.
\end{equation}
Solving these coupled equations, the metric function $V(r)$ assumes the form:
\begin{equation}\label{eq:Vr_general}
V(r) =
\begin{cases}
1 - \dfrac{M}{r} - \dfrac{\Lambda r^2}{3 g^2(\varepsilon)} - \dfrac{2 \sqrt{2} Q^3_\epsilon}{r g^2(\varepsilon)} \ln \left( \dfrac{r}{l} \right), & p = \dfrac{3}{2}, \\[12pt]
1 - \dfrac{M}{r} - \dfrac{\Lambda r^2}{3 g^2(\varepsilon)} - \dfrac{(2p - 1)^2 (2)^{p - 1} Q^{2p}_\epsilon}{(2p - 3) g^2(\varepsilon)} r^{-\frac{2}{2p - 1}}, & \dfrac{1}{2} < p < \dfrac{3}{2},
\end{cases}
\end{equation}
where $M$ is an integration constant identified with the BH mass. For the special case $p = 1$, the metric function reduces to the familiar form:
\begin{equation}\label{eq:Vr_p1}
V(r) = 1 - \frac{M}{r} - \frac{\Lambda r^2}{3 g^2(\varepsilon)} + \frac{f^2(\varepsilon) Q^2}{r^2},
\end{equation}
which describes the R-N-AdS BH modified by gravity's rainbow effects \cite{m7}. The rainbow function $g(\varepsilon)$ appears in the cosmological and charge terms, effectively rescaling their contributions to the geometry, while $f(\varepsilon)$ modifies only the charge sector.

\begin{figure}[h!]
\begin{center}
\subfigure[]{
\includegraphics[height=7cm,width=10cm]{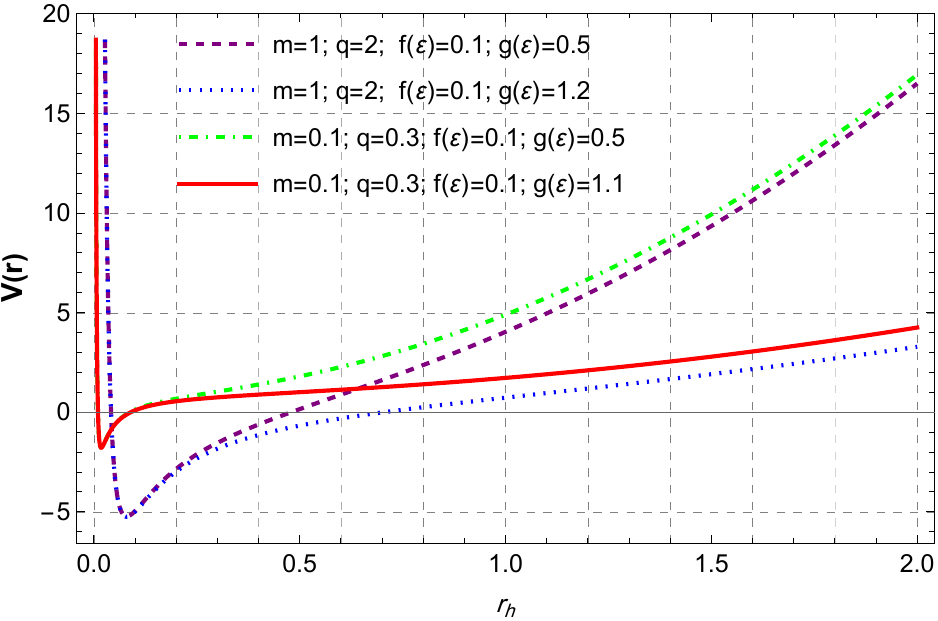}
\label{fig1a}}
\caption{The metric function $V(r)$ as a function of the radial coordinate for different parameter configurations. Four cases are shown: $(M=1, Q=2, f(\varepsilon)=0.1, g(\varepsilon)=0.5)$ depicted by the solid blue curve; $(M=1, Q=2, f(\varepsilon)=0.1, g(\varepsilon)=1.2)$ shown as the dashed red curve; $(M=0.1, Q=0.3, f(\varepsilon)=0.1, g(\varepsilon)=0.5)$ represented by the dash-dotted green curve; and $(M=0.1, Q=0.3, f(\varepsilon)=0.1, g(\varepsilon)=1.1)$ indicated by the dotted purple curve. All cases use $l=1$. The zero crossings mark the event horizon locations, demonstrating how rainbow parameters shift horizon radii relative to standard R-N-AdS BHs.}
\label{fig1}
\end{center}
\end{figure}

Figure~\ref{fig1} displays the behavior of the metric function $V(r)$ for several representative parameter choices. The plots reveal that horizon structure depends sensitively on both the rainbow parameters and the charge-to-mass ratio. When $g(\varepsilon) < 1$, the effective cosmological contribution strengthens, pushing the cosmological horizon inward and modifying the causal structure of the spacetime. Conversely, larger values of $g(\varepsilon)$ weaken these effects, approaching the standard GR limit.

The WGC occupies a fundamental position in the broader swampland program, which aims to distinguish effective field theories that can be consistently embedded in quantum gravity from those that cannot. As a central criterion, the WGC asserts that in any $U(1)$ gauge theory coupled to gravity, there must exist at least one particle whose charge-to-mass ratio satisfies $(q/m > 1)$, thereby ensuring that gravitational attraction never dominates over gauge forces. When this principle is examined in the context of black hole physics, it imposes significant restrictions on the allowable charge-to-mass ratios of black holes. Charged black holes with mass (M) and charge (Q) are commonly classified according to the relation between these parameters:
[
\text{subextremal: } $Q<M$,\quad
\text{extremal: } $Q= M$,\quad
\text{superextremal: } $Q> M$.
]
The WGC implies that extremal black holes should be unstable against emitting particles with sufficiently large charge-to-mass ratios, thereby preventing them from evolving into superextremal configurations. This mechanism protects the theory from producing naked singularities, maintaining consistency with the WCCC. Violation of the WGC would allow extremal black holes to transition into overcharged states, exposing spacetime singularities and conflicting with the foundational assumptions of classical general relativity. Thus, the presence of superextremal particles is essential for preserving the stability and evaporation processes of black holes, reinforcing the mutual compatibility of the WGC and WCCC.

Black hole thermodynamics provides a powerful framework for probing these ideas, offering insights into stability conditions, decay channels, and the compatibility of black hole solutions with quantum-gravity-motivated conjectures. Through this approach, one may examine both theoretical predictions and observational prospects, thereby strengthening the conceptual links between high-energy physics, gravitational theory, and cosmology.

To further analyze the mutual constraints posed by the WGC and WCCC, we consider the metric of Reissner–Nordström–Anti-de Sitter black holes modified by rainbow gravity, with particular emphasis on their horizon structure and physical characteristics. The location of the event horizon follows from solving (f(r)=0), where (M) and (Q) denote the mass and charge of the black hole. In the Reissner–Nordström black hole, the condition $(Q > M)$ immediately eliminates the event horizon, exposing the curvature singularity and thereby violating the WCCC, which requires all singularities to be concealed by horizons. By jointly solving the horizon equation together with the extremality condition—obtained either from the vanishing of the surface gravity or from the derivative of the metric function—we determine the precise circumstances under which both conjectures can be simultaneously satisfied. Because the WGC and WCCC are not automatically compatible across all black hole configurations, our analysis focuses on cases where the rainbow gravity plays a crucial role in reconciling these two principles. By isolating the contributions of the rainbow gravity parameter, we identify regimes in which extremal solutions satisfy both conjectures. Using standard WGC criteria, we then compute the extremal charge-to-mass boundary and delineate the parameter domain in which the conjecture remains valid.

Due to the algebraic complexity of the resulting equations—particularly those involving higher-order corrections—analytical solutions are generally unattainable, requiring numerical methods to chart the regions where the WGC and WCCC coexist. If such a region exists at extremality or near-critical configurations, this compatibility may extend to additional portions of the parameter space. By examining and categorizing the resulting classes of black holes, we identify promising candidates for testing ideas related to quantum gravity and the swampland program. This systematic study contributes to the broader effort of connecting black hole physics to foundational principles in high-energy theory and cosmology. For an extended discussion and related developments. To determine the extremality condition, we compute the quantities $r_{\text{ext}}$ and $M_{\text{ext}}$ using Eq.~(19). Accordingly, we obtain
\begin{equation*}\label{eq1}
r_{ext}=\sqrt{\frac{g^2(\varepsilon)}{2\Lambda}\Big[1\pm\sqrt{1-4\frac{\Lambda}{g^2(\varepsilon)}f^2(\varepsilon)Q^2}\Big]}
\end{equation*}
By substituting $r_{\text{ext}}$ into the mass relation and evaluating the extremality condition
$
\frac{q^{2}}{m^{2}}\ge \left(\frac{Q^{2}}{M^{2}}\right)_{\text{ext}}$, we obtain the bound
\begin{equation*}\label{eq2}
\begin{split}
M_{ext} =& (\sqrt{\frac{g^2(\varepsilon)}{2\Lambda}\Big[1\pm\sqrt{1-4\frac{\Lambda}{g^2(\varepsilon)}f^2(\varepsilon)Q^2}\Big]}) - \frac{\Lambda (\sqrt{\frac{g^2(\varepsilon)}{2\Lambda}\Big[1\pm\sqrt{1-4\frac{\Lambda}{g^2(\varepsilon)}f^2(\varepsilon)Q^2}\Big]})^3}{3 g^2(\varepsilon)} \\&+ \frac{f^2(\varepsilon) Q^2}{(\sqrt{\frac{g^2(\varepsilon)}{2\Lambda}\Big[1\pm\sqrt{1-4\frac{\Lambda}{g^2(\varepsilon)}f^2(\varepsilon)Q^2}\Big]})}
\end{split}
\end{equation*}

\subsection{PSs and WGC Compatibility}

The existence and properties of PSs provide crucial information about BH spacetimes and offer a natural connection to the WGC \cite{45'm,45mmm,45m,46m,47m,48m,49m,50m}. Following the topological approach developed in recent literature, we analyze the circular photon orbits in the equatorial plane ($\theta = \pi/2$), exploiting the $Z_2$ reflection symmetry of the spacetime.

The radial motion of photons obeys an energy-conservation-like equation:
\begin{equation}\label{p2}
\dot{r}^2 + V_{\text{eff}}(r) = 0,
\end{equation}
where the overdot denotes differentiation with respect to an affine parameter along the null geodesic. The effective potential governing photon trajectories takes the form:
\begin{equation}\label{p3}
V_{\text{eff}}(r) = g(r) \left(\frac{L^2}{r^2} - \frac{E_p^2}{V(r)} \right),
\end{equation}
with $E_p$ and $L$ being the conserved energy and angular momentum arising from the timelike and rotational Killing vectors, respectively. Circular photon orbits at radius $r_{\text{ps}}$ require:
\begin{equation}\label{p4}
V_{\text{eff}}(r_{\text{ps}}) = 0, \quad \text{and} \quad \frac{dV_{\text{eff}}}{dr}\Big|_{r=r_{\text{ps}}} = 0.
\end{equation}
These conditions combine to yield:
\begin{equation}\label{p5}
\frac{d}{dr}\left(\frac{V(r)}{r^2}\right)\Bigg|_{r=r_{\text{ps}}} = 0,
\end{equation}
which implicitly determines the PS radius. An equivalent formulation is:
\begin{equation}\label{p6}
2 r V(r) - r^2 V'(r) = 0.
\end{equation}
Stability analysis reveals that negative values of $d^2 V_{\text{eff}}/dr^2$ at $r_{\text{ps}}$ correspond to unstable orbits, while positive values indicate stability. For non-extremal BHs, PSs necessarily lie outside the event horizon since $V(r_h) = 0$ while $V'(r_h) \neq 0$ generically. Only in the extremal limit, where both conditions vanish simultaneously, can the PS radius coincide with the degenerate horizon \cite{45'm,45mmm,45m,46m,47m,48m,49m,50m}.

The topological classification of PSs employs the scalar function \cite{45'm,45mmm,45m,46m,47m,48m,49m,50m}:
\begin{equation}\label{p7}
H(r,\theta) = \sqrt{\frac{-g_{tt}}{g_{\phi\phi}}} = \frac{1}{\sin\theta} \sqrt{\frac{V(r)}{h(r)}},
\end{equation}
where $h(r)$ is an auxiliary radial function from the metric. The PS locations correspond to critical points satisfying:
\begin{equation}\label{p8}
\frac{dH}{dr} = 0.
\end{equation}
To construct the topological framework, we define a vector field $\boldsymbol{\varphi}$ in the $(r, \theta)$ plane with components:
\begin{equation}\label{p9}
\varphi^{r} = \sqrt{g(r)} \frac{dH}{dr}, \quad \varphi^\theta = \frac{\partial H/\partial \theta}{\sqrt{h(r)}}.
\end{equation}
This vector admits the polar representation:
\begin{equation}\label{p10}
\varphi = |\varphi| e^{i\Theta} = \varphi^{r} + i \varphi^\theta,
\end{equation}
with magnitude:
\begin{equation}\label{p11}
|\varphi| = \sqrt{(\varphi^{r})^2 + (\varphi^\theta)^2}.
\end{equation}
The normalized unit vector field reads:
\begin{equation}\label{p12}
n^a = \frac{\varphi^a}{|\varphi|},
\end{equation}
where $a=1,2$ labels the components. The topological charge associated with each PS encodes its stability: $\omega = -1$ for unstable PSs and $\omega = +1$ for stable ones. For the R-N-AdS BH in gravity's rainbow, the explicit scalar field components are:
\begin{equation}\label{p13}
\phi^{r} = -\frac{\csc(\theta) \left(4 f^2(\epsilon) Q^2 + r (2r - 3M)\right)}{2 r^4},
\end{equation}
and
\begin{equation}\label{p14}
\phi^\Theta = -\frac{\cot(\theta) \csc(\theta) \sqrt{\frac{f^2(\epsilon) Q^2}{r^2} + \frac{r^2}{g^2(\epsilon) l^2} - \frac{M}{r} + 1}}{r^2}.
\end{equation}

\begin{figure}[h!]
\begin{center}
\subfigure[]{
\includegraphics[height=3.5cm,width=3.5cm]{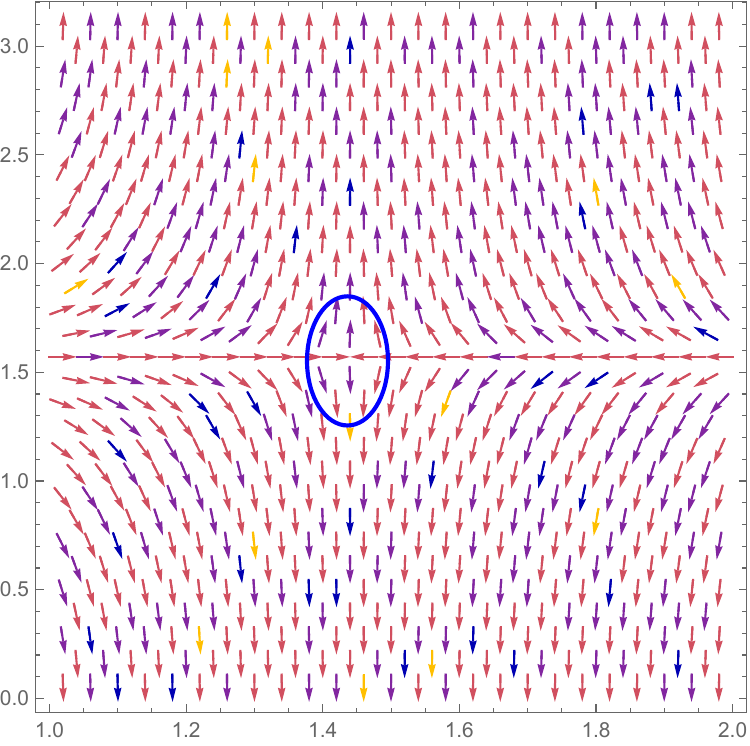}
\includegraphics[height=3.5cm,width=3.5cm]{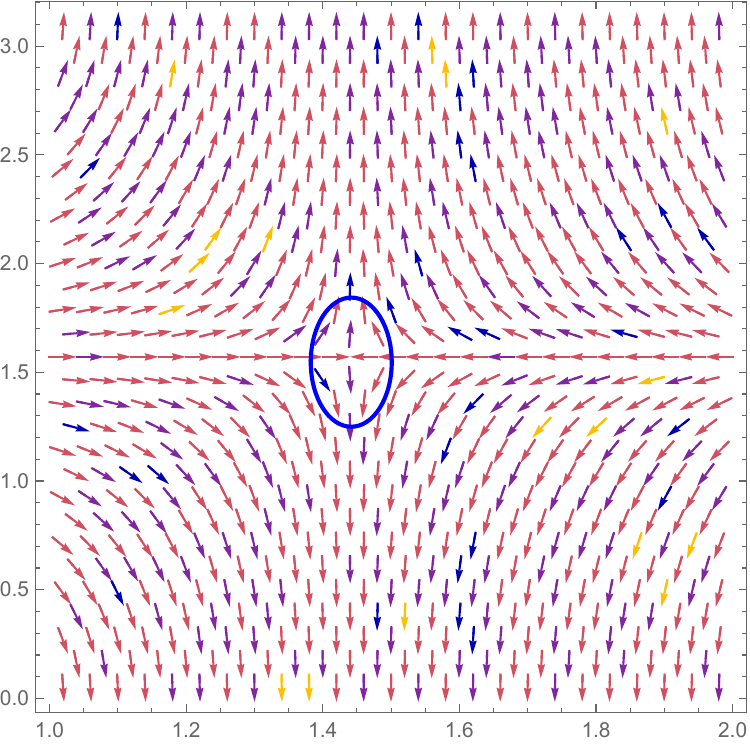}
\label{fig2a}}
\subfigure[]{
\includegraphics[height=3.5cm,width=3.5cm]{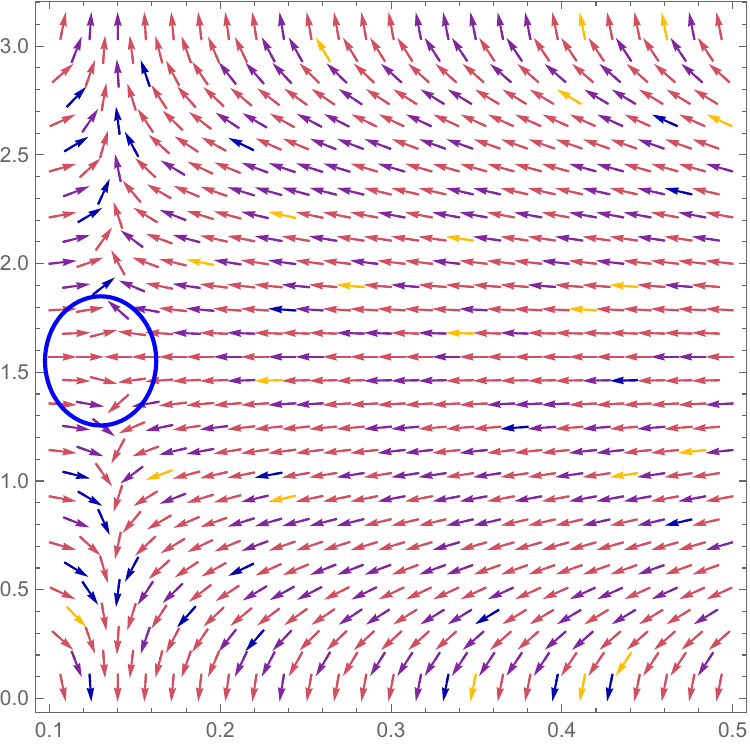}
\includegraphics[height=3.5cm,width=3.5cm]{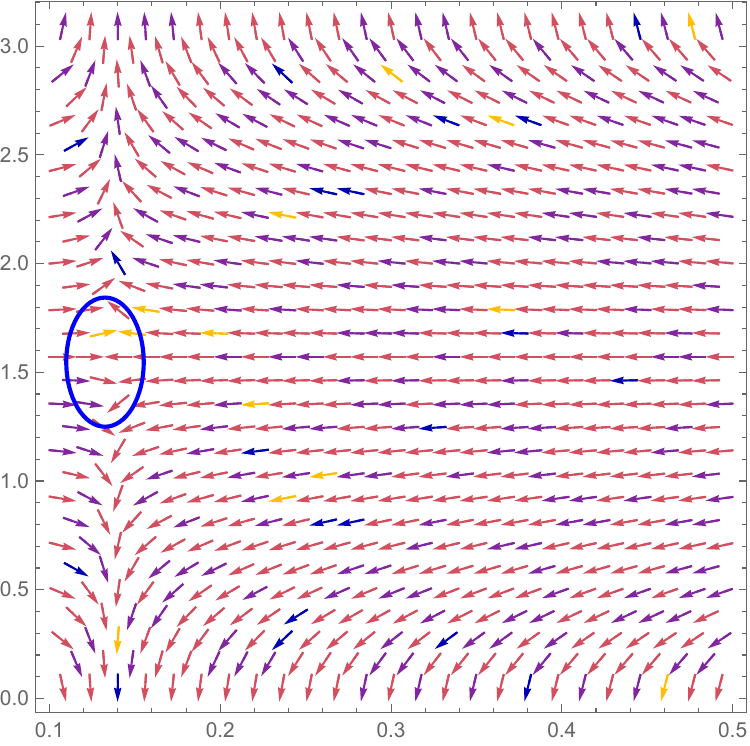}
\label{fig2b}}
\caption{Normalized vector field $\mathbf{n}$ in the $(r, \theta)$ plane showing PS locations for $l=1$ and $f(\epsilon)=0.1$. Panel (a): $Q=2$, $M=1$ with $g(\epsilon)=0.5$ (left) and $g(\epsilon)=1.2$ (right). Panel (b): $Q=0.3$, $M=0.1$ with $g(\epsilon)=0.5$ (left) and $g(\epsilon)=1.1$ (right). The zero points of the vector field (marked by circular patterns) indicate unstable PSs with topological charge $\omega=-1$. In all configurations, the PS resides outside the event horizon, validating both WGC and WCCC compatibility.}
\label{fig2}
\end{center}
\end{figure}

Figure~\ref{fig2} presents the normalized vector field in the $(r, \theta)$ plane for the parameter configurations listed in Table~\ref{tab:PS-WGC}. The vector flow patterns clearly exhibit zero points corresponding to unstable PSs with topological charge $\omega = -1$. These PSs are located outside the event horizons in all cases examined, confirming that the BH nature of the spacetime is preserved across the parameter space.

\begin{table}[h!]
\centering
\renewcommand{\arraystretch}{1.4}
\setlength{\tabcolsep}{12pt}
\begin{tabular}{ccccc}
\toprule
\textbf{$f(\epsilon)$} & \textbf{$g(\epsilon)$} & \textbf{PS} & \textbf{$q/m > (Q/M)_{\text{ext}}$} & \textbf{PS--WGC} \\
\midrule
0.1 & 0.4 & $-1$ & $\checkmark$ & $\checkmark$ \\
\rowcolor[HTML]{F5F5F5} 
0.1 & 0.5 & $-1$ & $\checkmark$ & $\checkmark$ \\
0.1 & 1.1 & $-1$ & $\checkmark$ & $\checkmark$ \\
\rowcolor[HTML]{F5F5F5} 
0.1 & 1.2 & $-1$ & $\checkmark$ & $\checkmark$ \\
\bottomrule
\end{tabular}
\caption{Verification of PS--WGC consistency for the R-N-AdS BH in gravity's rainbow with $l=1$. The PS column indicates the topological charge ($-1$ denotes unstable). The fourth column confirms that the WGC bound $q/m > (Q/M)_{\text{ext}}$ is satisfied, and the final column shows that both PS existence and WGC hold simultaneously for all tested parameter combinations.}
\label{tab:PS-WGC}
\end{table}

The results summarized in Table~\ref{tab:PS-WGC} and illustrated in Fig.~\ref{fig2} establish that unstable PSs exist for all examined rainbow parameter values. This finding carries significant implications: the presence of an unstable PS outside the horizon signals that the spacetime retains its BH character and does not develop naked singularities. From the WGC perspective, the PS provides a geometric manifestation of the balance between gravitational attraction and electromagnetic repulsion. When this balance permits circular photon orbits, it indicates that gravity is not the dominant force—precisely the condition demanded by the WGC.

Moreover, the simultaneous satisfaction of the WGC bound $q/m > (Q/M)_{\text{ext}}$ and the existence of PSs demonstrates that these rainbow-modified R-N-AdS BHs respect both the WGC and WCCC. The rainbow parameters shift the extremal charge-to-mass ratio as $(Q/M)_{\text{ext}} = 1/f(\varepsilon)$, which for $f(\varepsilon) = 0.1$ gives $(Q/M)_{\text{ext}} = 10$. This elevated threshold makes it easier for test particles to satisfy the WGC while keeping the BH sub-extremal, thereby protecting cosmic censorship. The gravity's rainbow framework thus provides a natural mechanism for reconciling these two fundamental conjectures in quantum gravity phenomenology.

\section{Gravitational Lensing Signatures and WGC Constraints at Extremality} \label{isec3}

Light propagation through curved spacetime remains one of the most effective observational tools for probing gravitational theories beyond GR. Gravitational lensing encodes rich information about the underlying geometry and matter-field interactions, offering direct observational signatures that can discriminate between competing theoretical models. In this section, we derive weak deflection angles for both photons and massive particles in the R-N-AdS BH geometry modified by gravity's rainbow using the GB theorem combined with the Jacobi-Maupertuis optical metric formalism. This geometric method, originally developed by Gibbons and Werner for static spacetimes and later extended to rotating geometries, recasts the deflection problem as an integration of Gaussian curvature over an appropriately constructed optical manifold \cite{039,040,041,042,043,044,045,sakalli_lens01,sakalli_lens02,sakalli_lens03,sakalli_lens04,sakalli_lens05,sakalli_lens06,sakalli_lens07}.

\subsection{Jacobi-Maupertuis Optical Geometry}

For the static, spherically symmetric spacetime given by Eq.~\eqref{m5}, test particle dynamics can be examined via the Jacobi-Maupertuis variational principle. A massive particle with rest mass $m$ and energy $E_p$ traces geodesics in a Riemannian optical geometry characterized by the Jacobi metric tensor:
\begin{equation}
\bar{\alpha}_{ij} = \frac{E_p^2 + m^2 V(r)}{V(r)} \gamma_{ij},
\label{jacobi_metric_rainbow}
\end{equation}
where $\gamma_{ij} = g_{ij} - g_{0i}g_{0j}/g_{00}$ represents the spatial metric sector that incorporates the rainbow functions. For our gravity's rainbow geometry, the spatial components become $\gamma_{rr} = g^2(\varepsilon)/V(r)$ and $\gamma_{\phi\phi} = g^2(\varepsilon)r^2$ when restricted to the equatorial plane. The energy $E_p$ connects to particle velocity $v$ via the relativistic expression $E_p = m/\sqrt{1-v^2}$ for massive particles. For photons, $E_p$ corresponds to the photon frequency, and the massless limit $m \to 0$ recovers the standard optical metric $\bar{\alpha}_{ij} = (E_p^2/V(r))\gamma_{ij}$.

Exploiting the spherical symmetry and confining the analysis to the equatorial plane $\theta = \pi/2$, the Jacobi metric components reduce to:
\begin{equation}
\bar{\alpha}_{rr} = \frac{g^2(\varepsilon)[E_p^2 + m^2 V(r)]}{V(r)^2}, \quad \bar{\alpha}_{\phi\phi} = \frac{g^2(\varepsilon)[E_p^2 + m^2 V(r)]}{V(r)} r^2.
\label{jacobi_components_rainbow}
\end{equation}
The metric determinant, which governs the integration measure when applying the GB theorem, takes the form:
\begin{equation}
\det\bar{\alpha} = \frac{g^4(\varepsilon)[E_p^2 + m^2 V(r)]^2 r^2}{V(r)^3},
\label{metric_determinant}
\end{equation}
encoding the rainbow modifications to the optical geometry volume element.

\subsection{Gaussian Curvature of the Optical Manifold}

The Gaussian curvature $\mathcal{K}$ of this two-dimensional Riemannian manifold dictates how light and particle trajectories bend as they traverse the spacetime. Following the Gibbons-Werner prescription, the Gaussian curvature is computed from the metric determinant and Christoffel symbols using standard differential geometry. The relevant nonvanishing connection coefficient is:
\begin{equation}
\Gamma^\phi_{rr} = -\frac{r V'(r)}{2V(r)[E_p^2 + m^2 V(r)]},
\label{christoffel_rainbow}
\end{equation}
where primes denote radial derivatives. For the R-N-AdS metric function with gravity's rainbow corrections:
\begin{equation}
V(r) = 1 - \frac{M}{r} + \frac{f^2(\varepsilon)Q^2}{r^2} + \frac{r^2}{g^2(\varepsilon)l^2},
\label{metric_function_full}
\end{equation}
with $l$ being the AdS radius, the Gaussian curvature assumes the explicit form:
\begin{equation}
\mathcal{K}(r) = \frac{V(r) V''(r) - 2[V'(r)]^2}{4g^2(\varepsilon)[E_p^2 + m^2 V(r)]^2} - \frac{V'(r)}{rg^2(\varepsilon)[E_p^2 + m^2 V(r)]} + \frac{m^2 [V'(r)]^2}{2g^2(\varepsilon)[E_p^2 + m^2 V(r)]^2}.
\label{gaussian_curvature_rainbow}
\end{equation}

The metric function derivatives required for the curvature calculation are:
\begin{equation}
V'(r) = \frac{M}{r^2} - \frac{2f^2(\varepsilon)Q^2}{r^3} + \frac{2r}{g^2(\varepsilon)l^2},
\label{V_prime}
\end{equation}
\begin{equation}
V''(r) = -\frac{2M}{r^3} + \frac{6f^2(\varepsilon)Q^2}{r^4} + \frac{2}{g^2(\varepsilon)l^2}.
\label{V_double_prime}
\end{equation}

\begin{figure}[h!]
\centering
\includegraphics[width=0.75\textwidth]{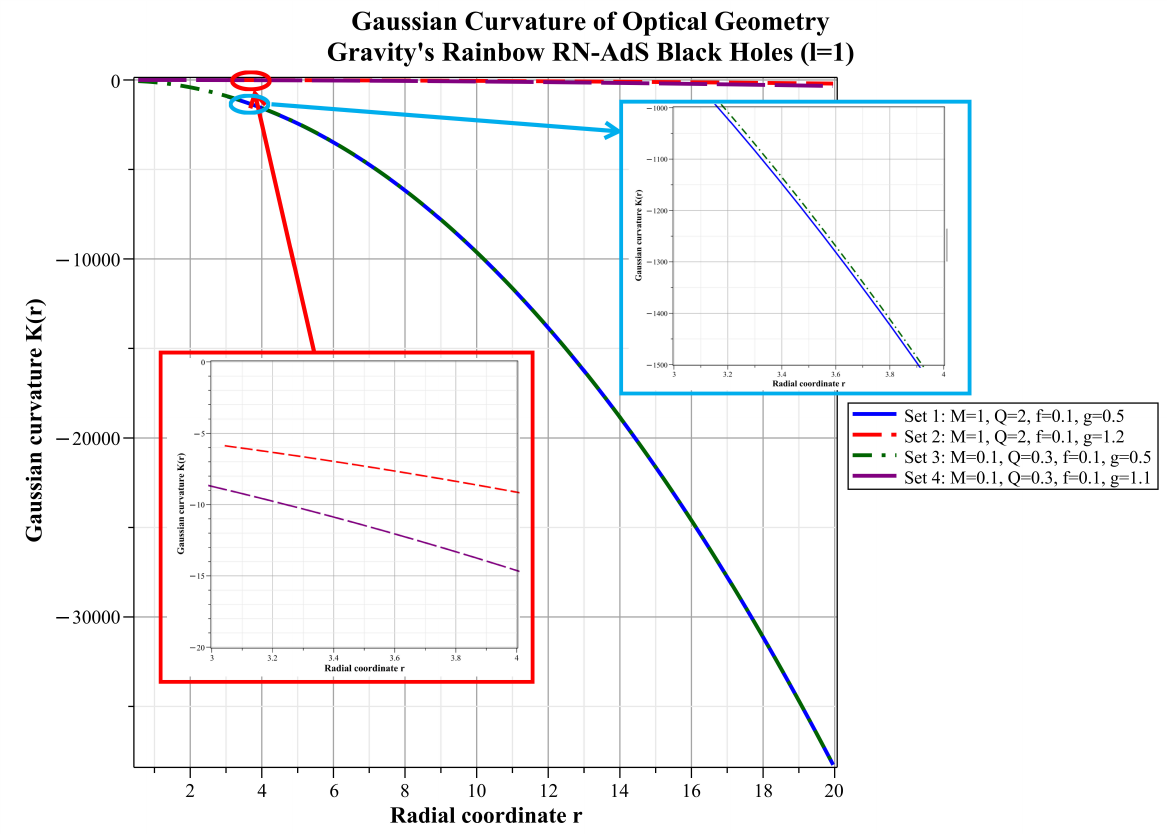}
\caption{Gaussian curvature $\mathcal{K}(r)$ of the Jacobi optical geometry for gravity's rainbow R-N-AdS BHs with AdS radius $l=1$. Four parameter sets are shown: Set 1 (blue solid): $M=1$, $Q=2$, $f=0.1$, $g=0.5$; Set 2 (red dashed): $M=1$, $Q=2$, $f=0.1$, $g=1.2$; Set 3 (green dash-dot): $M=0.1$, $Q=0.3$, $f=0.1$, $g=0.5$; Set 4 (purple solid): $M=0.1$, $Q=0.3$, $f=0.1$, $g=1.1$. The inset panels magnify the curvature structure near $r \sim 3$--$4$: the cyan inset (upper right) highlights Sets 1 and 3 with $g=0.5$, while the red inset (lower left) compares Sets 2 and 4 with $g > 1$. Sets 1 and 3 exhibit strongly negative curvature (reaching $\mathcal{K} \sim -35000$ at $r=20$), whereas Sets 2 and 4 show near-zero values. This dramatic difference quantifies the amplification induced by $g(\varepsilon) < 1$. The persistent negative curvature reflects the defocusing character of the optical geometry in Lorentzian spacetimes, with the magnitude directly controlling deflection angle strength.}
\label{fig:gaussian_curvature}
\end{figure}

The curvature profile captures how the rainbow parameters $f(\varepsilon)$ and $g(\varepsilon)$, along with the BH charge $Q$ and mass $M$, alter the optical geometry experienced by propagating photons. Figure~\ref{fig:gaussian_curvature} displays $\mathcal{K}(r)$ as a function of radial coordinate for four parameter configurations, exhibiting the characteristic negative curvature responsible for light bending toward the gravitating source.

\subsection{Weak Deflection Angle Derivation}

The weak deflection angle $\hat{\alpha}$ for asymptotically located source and observer is obtained by integrating the Gaussian curvature over the optical geometry exterior to the closest approach distance. The GB theorem, applied to the domain bounded by the particle trajectory and an asymptotic circle at infinity, yields:
\begin{equation}
\hat{\alpha} = -\int_0^{\pi} \int_{b/\sin\phi}^{\infty} \mathcal{K}(r) \sqrt{\det \bar{\alpha}} \, dr \, d\phi,
\label{deflection_integral_rainbow}
\end{equation}
where $b$ is the impact parameter and $r_0(\phi) = b/\sin\phi$ represents the unperturbed straight-line trajectory, valid in the weak-field regime $b \gg M, Q$.

Expanding the metric function under weak-field conditions ($r \gg M, Q$), where the cosmological constant contribution becomes negligible at finite distances, the first-order Gaussian curvature simplifies to:
\begin{equation}
\mathcal{K}^{(1)}(r) = \frac{M(1+v^2)}{g^2(\varepsilon)r^3 v^2} - \frac{2f^2(\varepsilon)Q^2(1+v^2)}{g^2(\varepsilon)r^4 v^2} + \mathcal{O}(r^{-5}),
\label{K_first_order_rainbow}
\end{equation}
using $E_p^2 = m^2/(1-v^2)$ for massive particles. The first term describes mass-induced (attractive) curvature, while the second captures the charge contribution (repulsive for like-sign charges).

Performing the radial integrations:
\begin{equation}
\int_{b/\sin\phi}^{\infty} \frac{dr}{r^2} = \frac{\sin\phi}{b}, \quad \int_{b/\sin\phi}^{\infty} \frac{dr}{r^3} = \frac{\sin^2\phi}{2b^2},
\label{radial_integrals}
\end{equation}
and the angular integrations $\int_0^{\pi} \sin\phi \, d\phi = 2$ and $\int_0^{\pi} \sin^2\phi \, d\phi = \pi/2$, we obtain the first-order deflection:
\begin{equation}
\hat{\alpha}^{(1)} = \frac{4M(1+v^2)}{g^2(\varepsilon)bv^2} - \frac{\pi f^2(\varepsilon)Q^2(1+v^2)}{g^2(\varepsilon)b^2 v^2}.
\label{alpha_first_rainbow}
\end{equation}

The second-order Schwarzschild correction, arising from higher-order curvature terms and quadratic mass contributions, gives:
\begin{equation}
\hat{\alpha}^{(2)}_{\rm Schw} = \frac{3\pi M^2(4+v^2)}{4g^4(\varepsilon)b^2v^4}.
\label{alpha_second_schw_rainbow}
\end{equation}

The complete weak deflection angle to second order reads:
\begin{equation}
\hat{\alpha} = \frac{4M(1+v^2)}{g^2(\varepsilon)bv^2} - \frac{\pi f^2(\varepsilon)Q^2(1+v^2)}{g^2(\varepsilon)b^2 v^2} + \frac{3\pi M^2(4+v^2)}{4g^4(\varepsilon)b^2v^4} + \mathcal{O}(M^3/b^3).
\label{deflection_complete_rainbow}
\end{equation}

For photons ($v = 1$), this reduces to:
\begin{equation}
\hat{\alpha}_{\rm photon} = \frac{4M}{g^2(\varepsilon)b} - \frac{2\pi f^2(\varepsilon)Q^2}{g^2(\varepsilon)b^2} + \frac{15\pi M^2}{4g^4(\varepsilon)b^2} + \mathcal{O}(M^3/b^3).
\label{deflection_photon_rainbow}
\end{equation}

\begin{figure}[h!]
\centering
\includegraphics[width=0.825\textwidth]{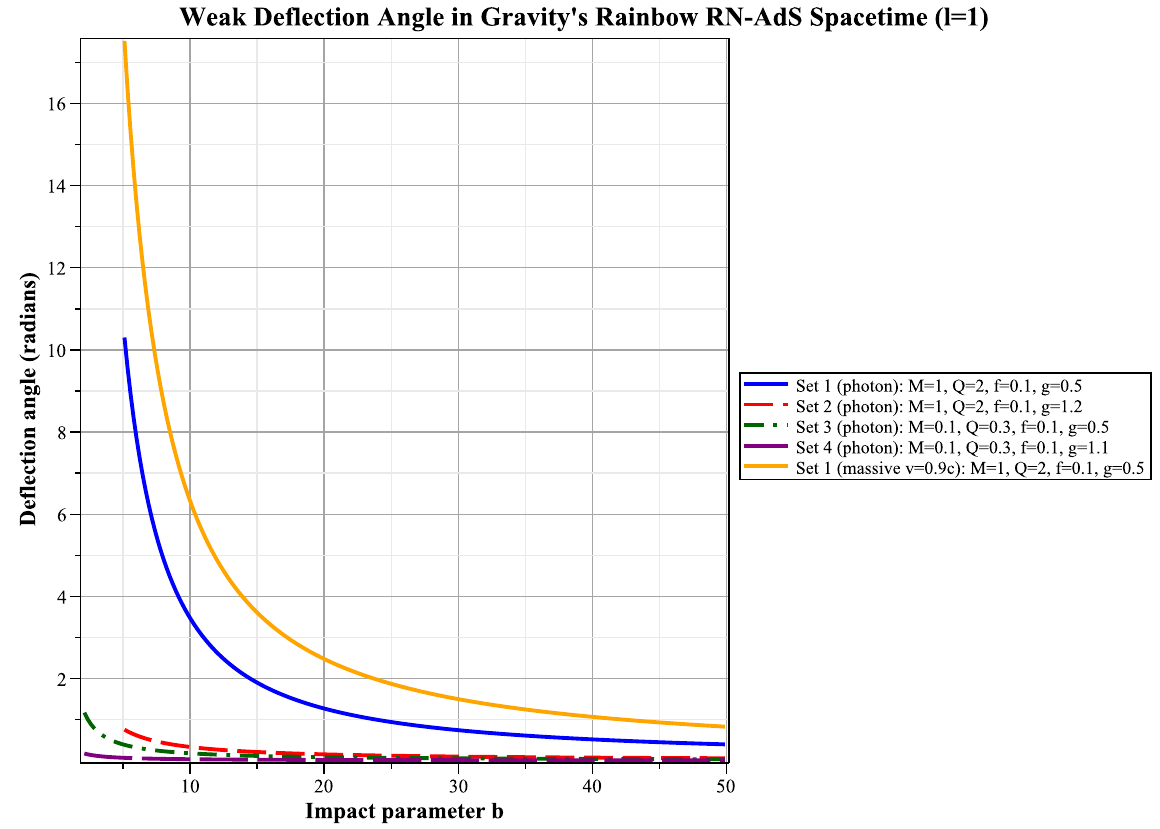}
\caption{Weak deflection angle $\hat{\alpha}$ versus impact parameter $b$ for gravity's rainbow R-N-AdS BHs with $l=1$. Five configurations are displayed: Set 1 photon (blue solid): $M=1$, $Q=2$, $f=0.1$, $g=0.5$; Set 2 photon (red dashed): $M=1$, $Q=2$, $f=0.1$, $g=1.2$; Set 3 photon (green dash-dot): $M=0.1$, $Q=0.3$, $f=0.1$, $g=0.5$; Set 4 photon (purple long-dash): $M=0.1$, $Q=0.3$, $f=0.1$, $g=1.1$; Set 1 massive $v=0.9c$ (orange solid): $M=1$, $Q=2$, $f=0.1$, $g=0.5$. All curves follow the characteristic $\hat{\alpha} \propto 1/b$ scaling at large $b$, confirming first-order dominance. The marked enhancement for $g=0.5$ (Sets 1, 3) relative to $g > 1$ (Sets 2, 4) demonstrates the $g^{-2}$ amplification. The massive particle curve (orange) exceeds its photon counterpart due to the velocity factor $(1+v^2)/v^2 \approx 2.23$ for $v=0.9c$.}
\label{fig:deflection_main}
\end{figure}

The rainbow functions enter the deflection formula in distinct ways: $g(\varepsilon)$ appears in denominators with powers $g^{-2}$ and $g^{-4}$, amplifying deflection when $g(\varepsilon) < 1$, while $f(\varepsilon)$ enters only through the charge term as $f^2$, modulating the electromagnetic contribution to light bending. Figure~\ref{fig:deflection_main} illustrates these dependencies for the primary parameter configurations.

\begin{figure}[h!]
\centering
\subfigure[]{\includegraphics[width=0.48\textwidth]{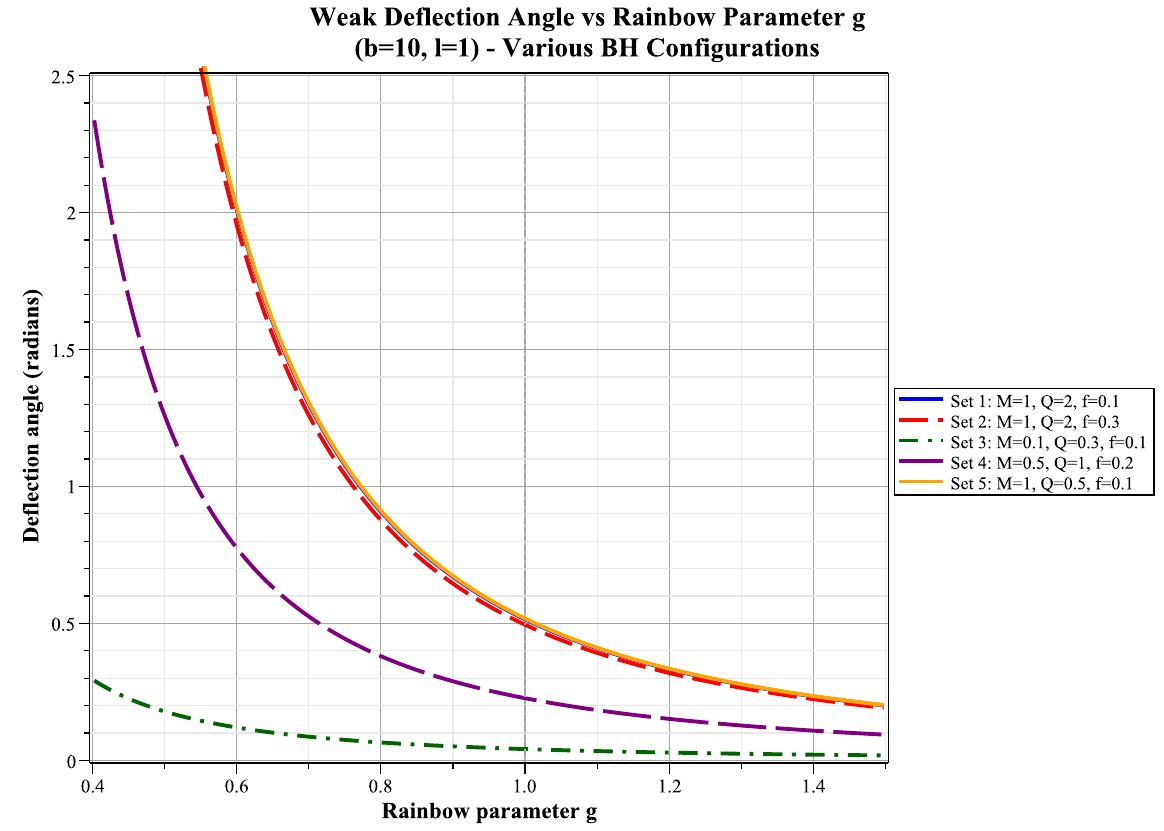}\label{fig:extra_g}}
\subfigure[]{\includegraphics[width=0.48\textwidth]{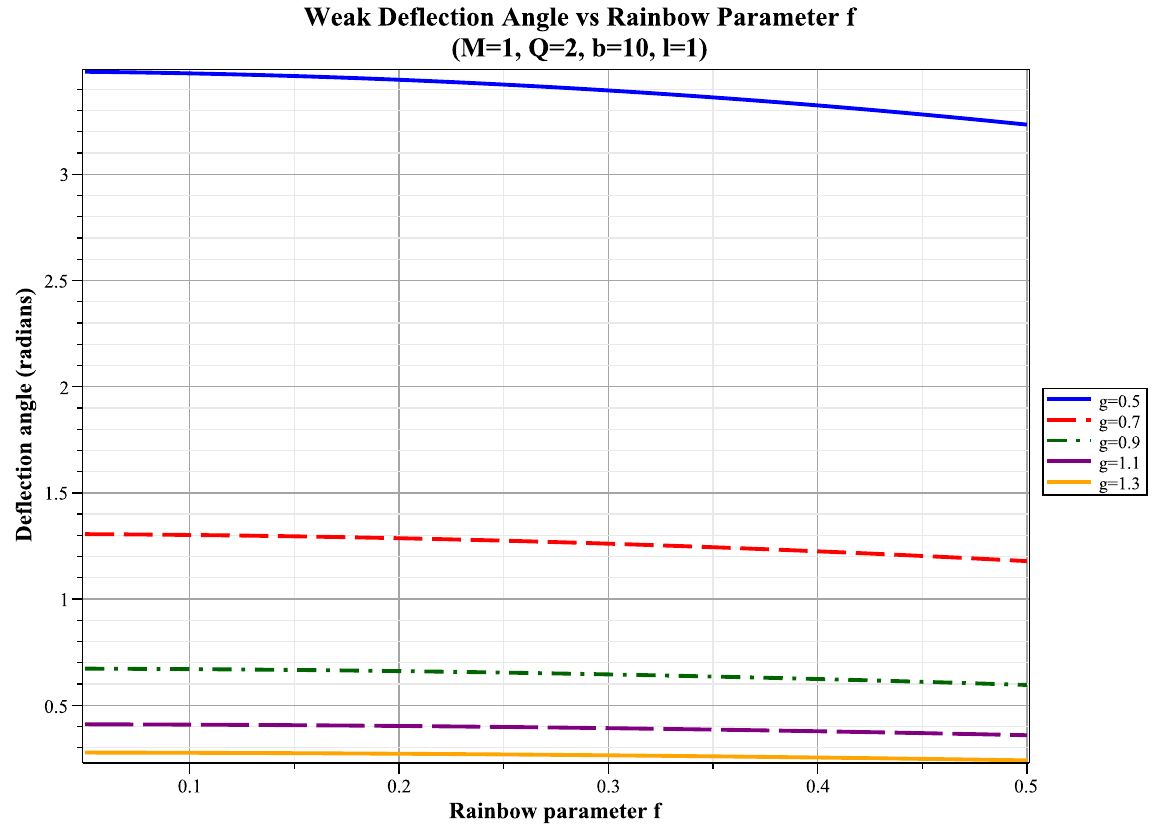}\label{fig:extra_massive}}
\subfigure[]{\includegraphics[width=0.48\textwidth]{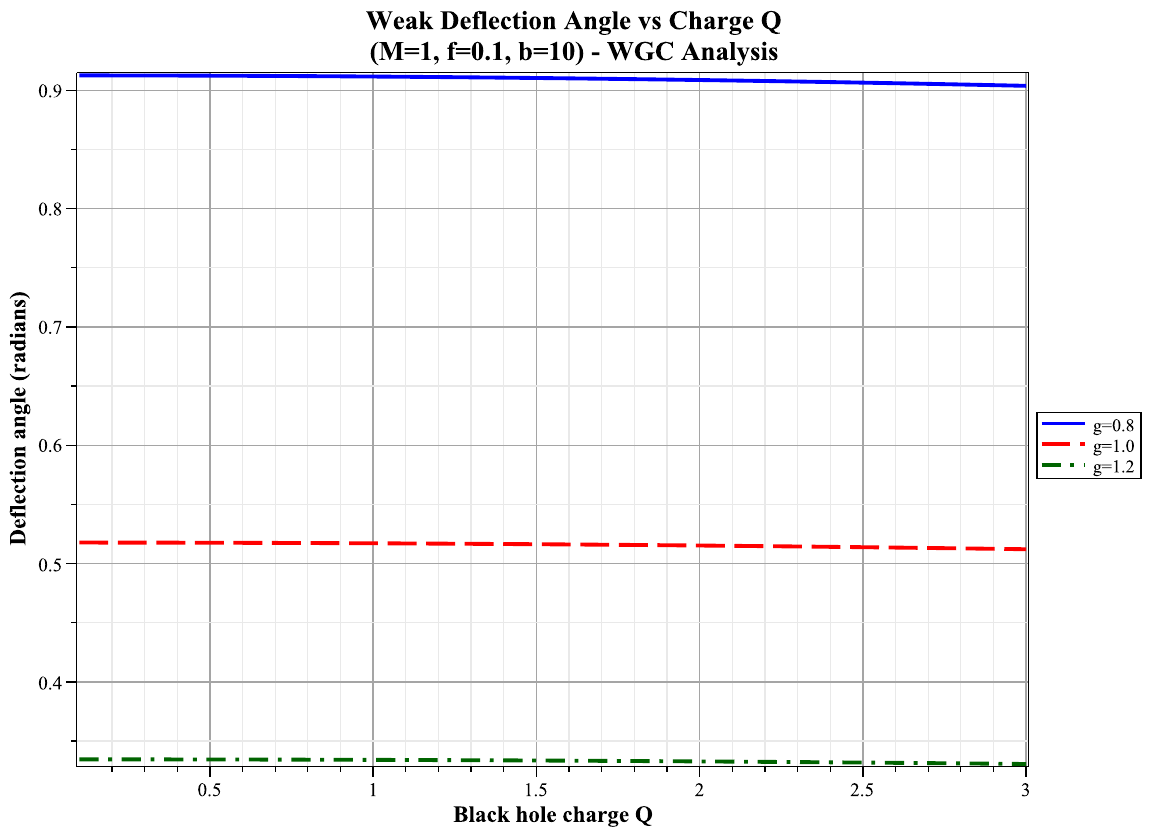}\label{fig:extra_charge}}
\subfigure[]{\includegraphics[width=0.48\textwidth]{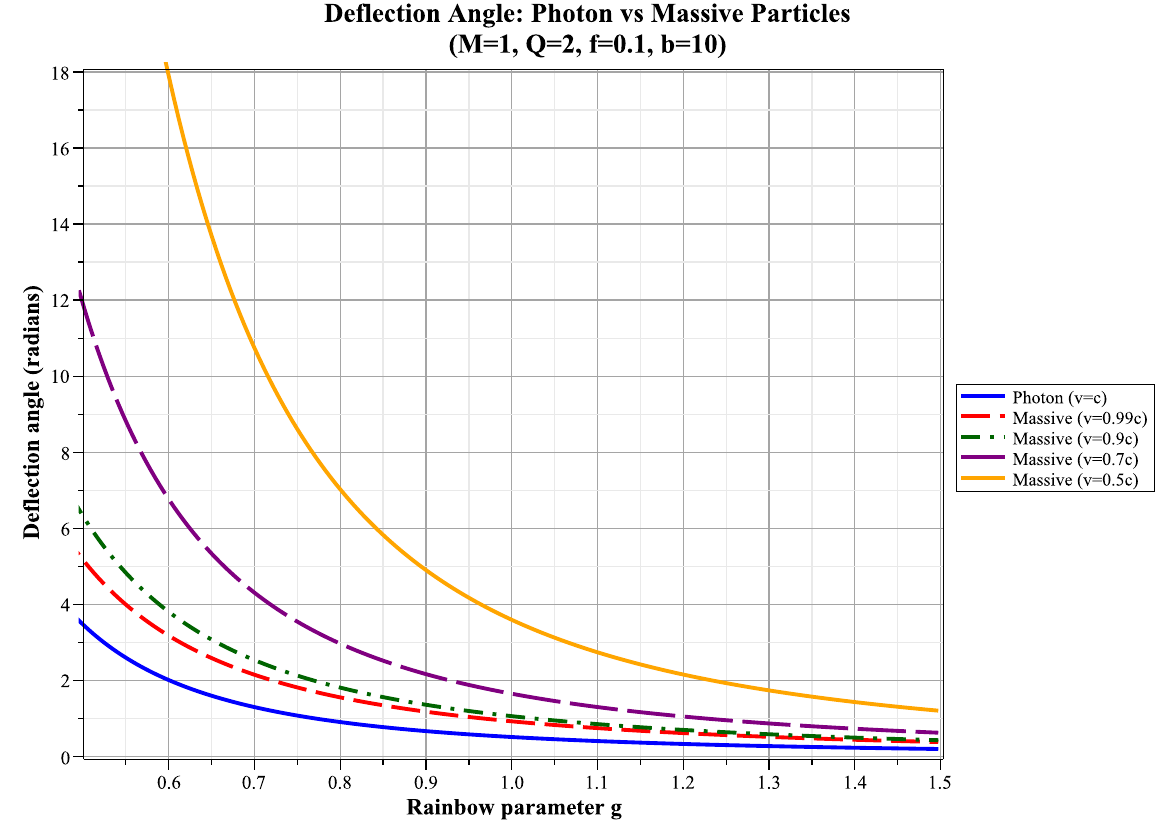}\label{fig:extra_extended}}
\caption{Extended deflection angle analysis for gravity's rainbow parameters. (a) Deflection versus $g(\varepsilon)$ at fixed $M$, $Q$, $f$, showing the $g^{-2}$ enhancement as $g$ decreases below unity. (b) Photon ($v=c$) versus massive particle ($v=0.5c, 0.7c, 0.9c, 0.99c$) deflection, revealing the velocity-dependent amplification $(1+v^2)/v^2$ that enhances bending for slower particles. (c) Charge variation effects, illustrating the negative (repulsive) $Q^2$ contribution that reduces net deflection for highly charged BHs. (d) Extended parameter space mapping combining variations in $M$, $Q$, and rainbow functions across the theory parameter space.}
\label{fig:deflection_extra}
\end{figure}

Figure~\ref{fig:deflection_extra} extends the analysis across broader parameter ranges, examining how charge magnitude, rainbow functions, and particle velocity affect the deflection signatures.

\subsection{WGC Implications at Extremality}

The extremal condition for R-N-AdS BHs in gravity's rainbow occurs when the horizon degenerates to a single root, marking the boundary between BH solutions (with event horizons) and naked singularities (without horizons). For the metric function $V(r)$, extremality demands $V(r_e) = 0$ and $V'(r_e) = 0$ simultaneously, where $r_e$ is the extremal horizon radius. In the asymptotically flat limit (neglecting the cosmological constant), these conditions yield the modified extremal charge-to-mass ratio,
$
(Q/M)_{\rm ext}\simeq 1/f(\varepsilon).
$
This result has deep implications for the WGC, which requires the existence of particles with $q/m \geq (Q/M)_{\rm ext}$ to ensure BH decay channels remain kinematically open, thereby preventing stable extremal remnants. When $f(\varepsilon) < 1$, the extremal ratio exceeds unity ($(Q/M)_{\rm ext} > 1$), effectively relaxing the WGC bound by allowing larger charge-to-mass ratios before extremality is reached. Conversely, $f(\varepsilon) > 1$ tightens the constraint, demanding particles with smaller $q/m$ values. This rainbow-induced shift provides a natural mechanism linking quantum gravity corrections to swampland constraints.

At extremality, with $Q_{\rm ext} = M/f(\varepsilon)$ the charge term simplifies since $f^2(\varepsilon)Q_{\rm ext}^2 = M^2$, yielding:
\begin{equation}
\hat{\alpha}_{\rm ext} = \frac{4M}{g^2(\varepsilon)b} - \frac{2\pi M^2}{g^2(\varepsilon)b^2} + \frac{15\pi M^2}{4g^4(\varepsilon)b^2}.
\label{deflection_extremal}
\end{equation}

Remarkably, this extremal deflection angle is \textit{independent} of $f(\varepsilon)$---only $g(\varepsilon)$ and $M$ determine lensing at the extremal bound. This $f$-independence emerges directly from the extremality condition: the product $f^2 Q_{\rm ext}^2 = M^2$ is fixed regardless of the specific $f(\varepsilon)$ value. This universal behavior offers a robust observational prediction that can be tested without prior knowledge of the rainbow function $f(\varepsilon)$.

\begin{figure}[h!]
\centering
\includegraphics[width=0.99\textwidth]{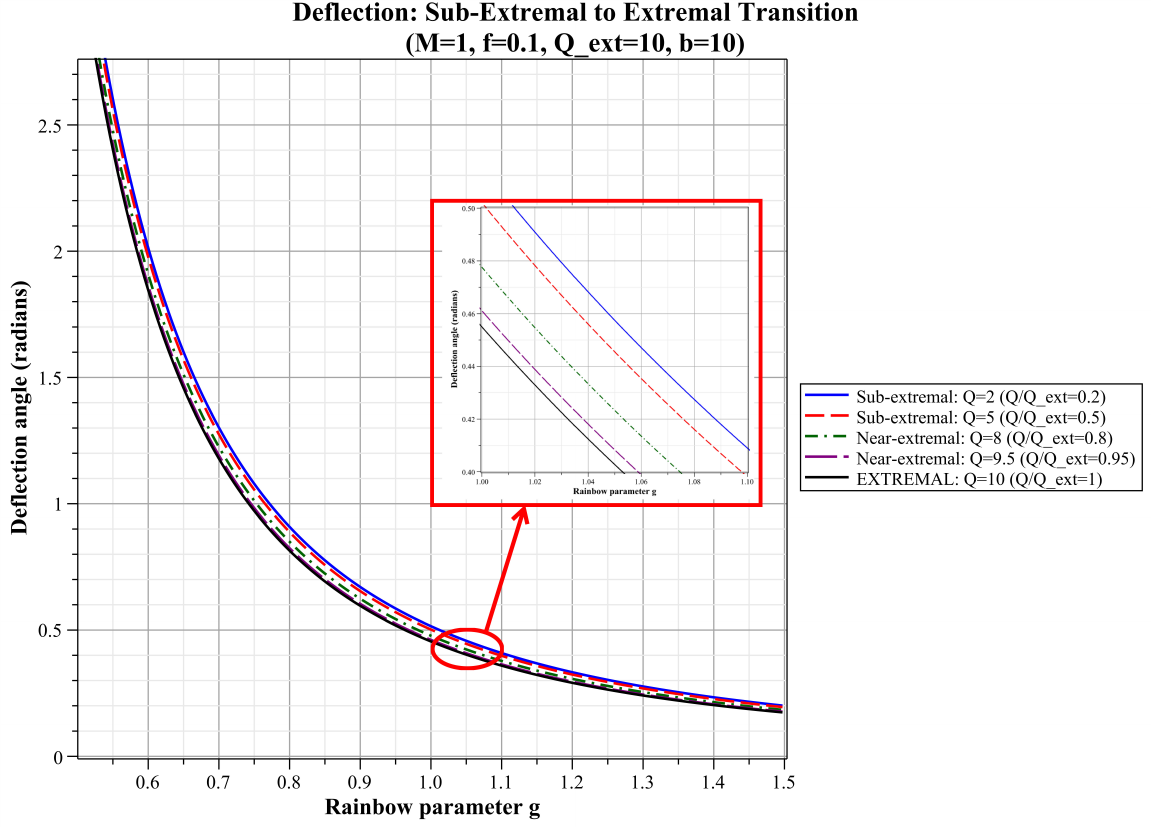}
\caption{Deflection angle evolution from sub-extremal to extremal configurations for gravity's rainbow R-N-AdS BHs with $M=1$, $f=0.1$, $b=10$, giving extremal charge $Q_{\rm ext}=M/f=10$. Five charge values are shown: $Q=2$ (blue solid, $Q/Q_{\rm ext}=0.2$), $Q=5$ (red dashed, $Q/Q_{\rm ext}=0.5$), $Q=8$ (green dash-dot, $Q/Q_{\rm ext}=0.8$), $Q=9.5$ (purple long-dash, $Q/Q_{\rm ext}=0.95$), and $Q=10$ (black solid, extremal). The inset magnifies the region near $g \sim 1$ where percent-level curve separations become visible. As charge approaches extremality, the negative $Q^2$ term in Eq.~\eqref{deflection_photon_rainbow} grows, reducing the net deflection. The extremal curve (black) represents the minimum deflection achievable for given $M$ and $g(\varepsilon)$, serving as a theoretical lower bound for sub-extremal BH lensing.}
\label{fig:extremal_transition}
\end{figure}

Figure~\ref{fig:extremal_transition} displays the transition from sub-extremal to extremal configurations, revealing how the deflection angle evolves as the BH approaches extremality.

\begin{figure}[h!]
\centering
\includegraphics[width=0.97\textwidth]{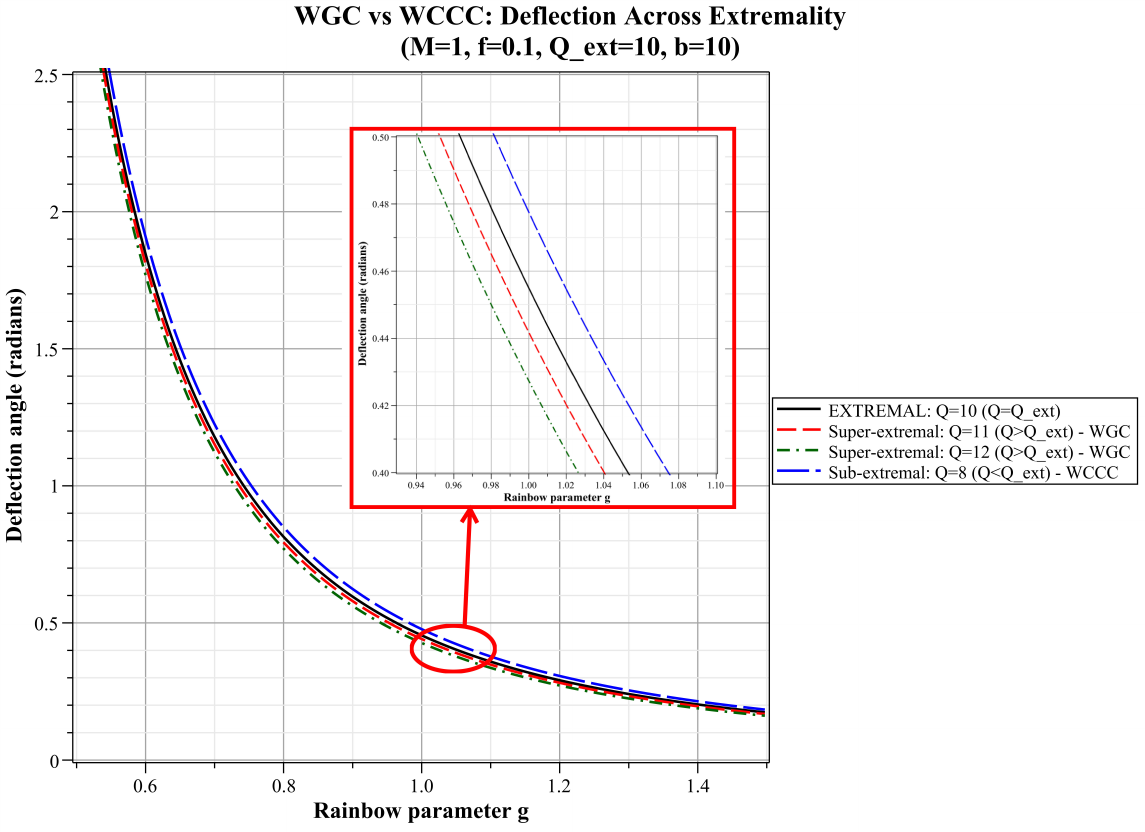}
\caption{WGC versus WCCC: deflection angle comparison across extremality for $M=1$, $f=0.1$, $Q_{\rm ext}=10$, $b=10$. Four configurations span the extremality boundary: sub-extremal $Q=8$ (blue long-dash, WCCC satisfied), extremal $Q=10$ (black solid, boundary case), and super-extremal $Q=11, 12$ (red dashed and green dash-dot, WGC satisfied but WCCC violated). The inset reveals the ordering reversal: super-extremal configurations exhibit \textit{larger} deflection angles than extremal ones due to the enhanced $Q^2$ contribution, despite being naked singularities rather than BHs. This counterintuitive result---WCCC-violating spacetimes producing stronger lensing---offers a potential observational discriminant. Detection of anomalously strong lensing from compact objects could indicate super-extremal configurations permitted by WGC, providing empirical tests of swampland conjectures.}
\label{fig:wgc_wccc}
\end{figure}

The competition between the WGC and WCCC becomes apparent when examining super-extremal configurations with $Q > Q_{\rm ext}$. Such configurations violate WCCC by exposing naked singularities, yet they naturally satisfy WGC by providing decay products with $q/m > (Q/M)_{\rm ext}$. Figure~\ref{fig:wgc_wccc} contrasts these scenarios through their lensing signatures.

\begin{figure}[h!]
\centering
\includegraphics[width=0.85\textwidth]{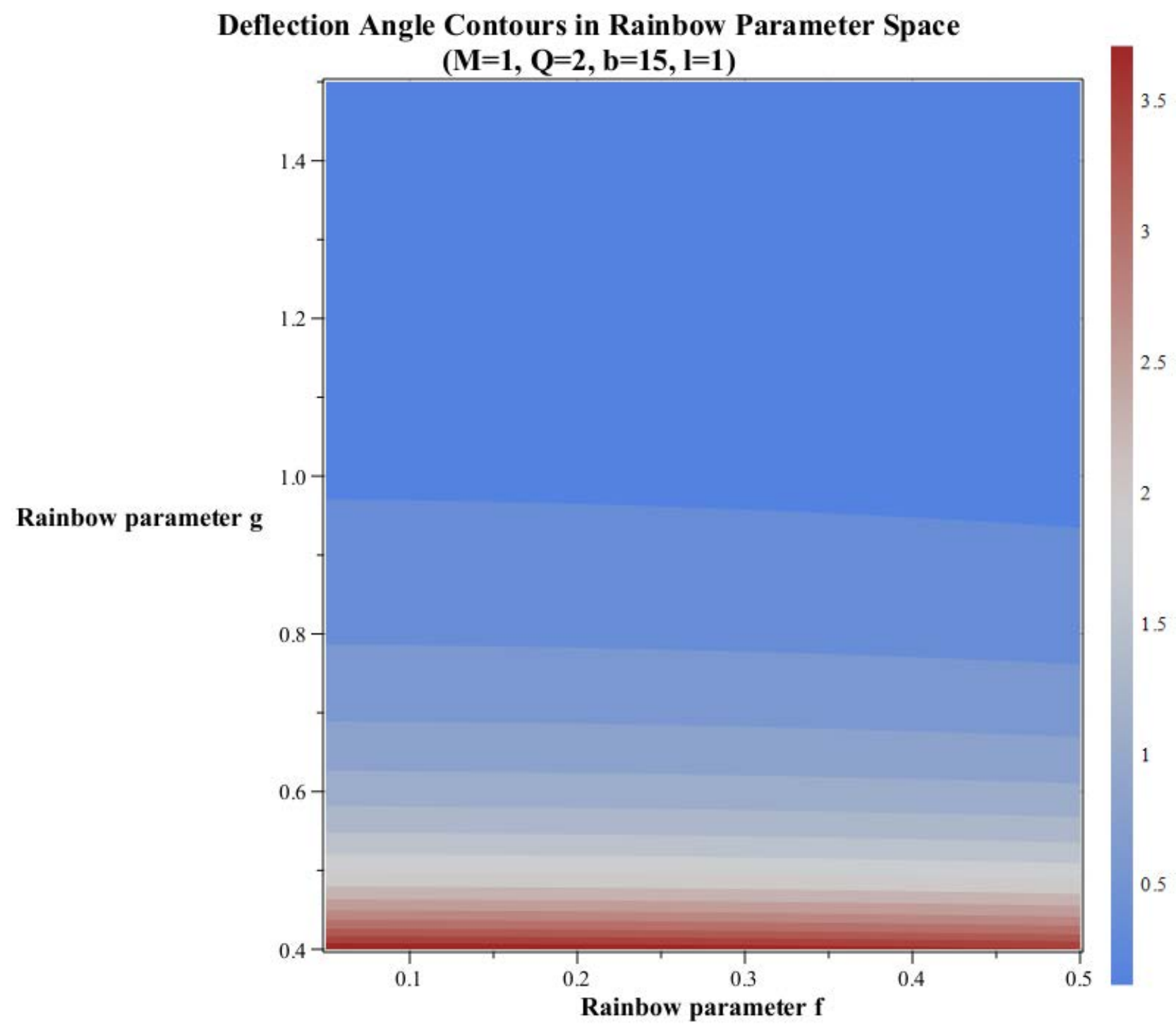}
\caption{Deflection angle contours in rainbow parameter space $(f, g)$ for $M=1$, $Q=2$, $b=15$, $l=1$. The color gradient from blue ($\hat{\alpha} \sim 0.3$ rad) to red ($\hat{\alpha} \sim 3.5$ rad) reveals the dominant role of $g(\varepsilon)$: deflection increases dramatically as $g \to 0.4$, achieving order-of-magnitude enhancement compared to the GR limit $g = 1$. The pronounced horizontal stratification confirms that $g(\varepsilon)$ controls the primary lensing amplification through the $g^{-2}$ and $g^{-4}$ factors in Eq.~\eqref{deflection_photon_rainbow}, while $f$-dependence remains subdominant due to the small $f^2 Q^2$ contribution at moderate charges. This hierarchy indicates that gravitational lensing observations are primarily sensitive to $g(\varepsilon)$, with $f(\varepsilon)$ constraints requiring either highly charged BHs or complementary observables such as the extremality bound in $
(Q/M)_{\rm ext}\simeq 1/f(\varepsilon).
$.}
\label{fig:contour}
\end{figure}

The two-dimensional rainbow parameter space $(f, g)$ provides a visualization of lensing modifications across the full parameter range. Figure~\ref{fig:contour} presents contour maps showing the dominant role of $g(\varepsilon)$ in controlling deflection strength.

\subsection{Observational Predictions and Parameter Constraints}

The preceding analysis establishes several key observational predictions for testing gravity's rainbow through gravitational lensing:

\begin{enumerate}
\item \textbf{Primary sensitivity to $g(\varepsilon)$}: The deflection angle scales as $g^{-2}$ at leading order, yielding strong sensitivity to this rainbow parameter. Precision astrometry of multiply-imaged quasars can constrain deviations from $g=1$ at the percent level.

\item \textbf{Extremality as universal probe}: The $f$-independence of extremal deflection (Eq.~\eqref{deflection_extremal}) implies that near-extremal BH lensing probes $g(\varepsilon)$ without degeneracy from $f(\varepsilon)$ uncertainty.

\item \textbf{WGC/WCCC discrimination}: Super-extremal configurations produce enhanced deflection compared to extremal BHs, offering a potential signature for WGC-satisfying naked singularities that would challenge WCCC.

\item \textbf{Velocity-dependent signatures}: Massive particle deflection exceeds photon deflection by the factor $(1+v^2)/v^2$, enabling complementary constraints through cosmic ray or neutrino lensing observations.
\end{enumerate}

These predictions, combined with upcoming observational facilities such as SKA, ngEHT, and space-based gravitational wave detectors, position gravitational lensing as a powerful probe of quantum gravity phenomenology through gravity's rainbow modifications to charged AdS BH spacetimes. The distinct $g^{-2}$ scaling and the universal extremal deflection formula provide concrete
\newpage
\section{Conclusion} \label{isec4}

In this work, we investigated the interplay between gravitational lensing, PSs, and the WGC within the framework of gravity's rainbow applied to R-N-AdS BHs. The primary motivation stemmed from the long-standing tension between the WGC and WCCC---two foundational conjectures in quantum gravity that impose seemingly contradictory constraints on charged BH configurations. By incorporating energy-dependent rainbow functions into the spacetime geometry, we explored how Planck-scale modifications alter both the extremality bounds and the observational signatures accessible through gravitational lensing.

In Section~\ref{isec2}, we constructed the R-N-AdS BH solution modified by gravity's rainbow effects, starting from the Einstein-NLED action given in Eq.~\eqref{m1}. The rainbow functions $f(\varepsilon)$ and $g(\varepsilon)$, which encode quantum gravity corrections through their dependence on the ratio $\varepsilon = E/E_P$, entered the metric structure in distinct ways: $f(\varepsilon)$ modified the temporal sector while $g(\varepsilon)$ rescaled spatial components. We derived the metric function $V(r)$ for general nonlinearity parameter $p$ (Eq.~\eqref{eq:Vr_general}) and examined the special case $p=1$ corresponding to linear Maxwell electrodynamics (Eq.~\eqref{eq:Vr_p1}). The horizon structure, illustrated in Fig.~\ref{fig1}, demonstrated that rainbow parameters significantly shift horizon locations relative to standard R-N-AdS BHs, with $g(\varepsilon) < 1$ strengthening the effective cosmological contribution.

The topological analysis of PSs presented in Section~\ref{isec2} employed the framework to classify circular photon orbits according to their stability properties. We derived the scalar field components $\phi^r$ and $\phi^\Theta$ (Eqs.~\eqref{p13}--\eqref{p14}) and constructed the normalized vector field whose zero points correspond to PS locations. As shown in Fig.~\ref{fig2} and summarized in Table~\ref{tab:PS-WGC}, all examined parameter configurations exhibited unstable PSs with topological charge $\omega = -1$ located outside the event horizon. This finding confirmed that the BH nature of the spacetime remained intact across the parameter space and that the WGC bound $q/m > (Q/M)_{\rm ext}$ was satisfied simultaneously with PS existence.

Section~\ref{isec3} developed the gravitational lensing formalism using the GB theorem combined with the Jacobi-Maupertuis optical geometry. The Gaussian curvature of the optical manifold (Eq.~\eqref{gaussian_curvature_rainbow}), displayed in Fig.~\ref{fig:gaussian_curvature}, revealed that rainbow parameters with $g(\varepsilon) < 1$ dramatically amplified the negative curvature responsible for light bending. We derived the complete weak deflection angle to second order (Eq.~\eqref{deflection_complete_rainbow}), which for photons reduced to Eq.~\eqref{deflection_photon_rainbow}. The rainbow function $g(\varepsilon)$ appeared with powers $g^{-2}$ and $g^{-4}$, providing strong observational sensitivity, while $f(\varepsilon)$ entered only through the charge-dependent term as $f^2$. Figure~\ref{fig:deflection_main} illustrated these dependencies, showing order-of-magnitude enhancements in deflection angle for $g(\varepsilon) = 0.5$ compared to the GR limit $g = 1$.

A central result of our analysis concerned the modified extremality condition $(Q/M)_{\rm ext} = 1/f(\varepsilon)$. This relation demonstrated that gravity's rainbow provides a natural mechanism for modulating the WGC bound: when $f(\varepsilon) < 1$, the extremal ratio exceeds unity, effectively relaxing the WGC constraint by permitting larger charge-to-mass ratios before the BH becomes extremal. At extremality itself, the deflection angle (Eq.~\eqref{deflection_extremal}) exhibited a remarkable property---complete independence from $f(\varepsilon)$. Only $g(\varepsilon)$ and the mass $M$ determined lensing at the extremal bound, offering a universal prediction testable without prior knowledge of the specific rainbow function $f(\varepsilon)$.

The competition between the WGC and WCCC was examined through the lensing signatures of sub-extremal, extremal, and super-extremal configurations, as displayed in Figs.~\ref{fig:extremal_transition} and \ref{fig:wgc_wccc}. We found that super-extremal configurations, which violate WCCC by exposing naked singularities, produce enhanced deflection angles compared to extremal BHs. This counterintuitive result---where WCCC-violating spacetimes generate stronger lensing---provides a potential observational discriminant between quantum gravity scenarios. The contour plot in Fig.~\ref{fig:contour} mapped the deflection angle across the $(f, g)$ parameter space, confirming the dominant role of $g(\varepsilon)$ through pronounced horizontal stratification.

Our findings connect to broader developments in the swampland program and quantum gravity phenomenology. The WGC, originally formulated to prevent stable extremal remnants that would conflict with BH thermodynamics and holography, has found applications ranging from axion physics to inflationary cosmology. Our work extended these considerations to the observational realm by establishing concrete lensing predictions that can discriminate between WGC-compatible and WCCC-compatible configurations. The velocity-dependent amplification factor $(1+v^2)/v^2$ in the massive particle deflection formula (Eq.~\eqref{deflection_complete_rainbow}) opens additional channels for testing gravity's rainbow through cosmic ray or neutrino lensing.

From an observational perspective, the predictions developed here align well with the capabilities of current and upcoming facilities. Precision astrometry with ngEHT can resolve deflection angles at the microarcsecond level, sufficient to detect percent-level deviations from the GR prediction $g = 1$. The SKA, with its unprecedented sensitivity to radio transients, offers complementary constraints through time-delay measurements in multiply-imaged systems. Gravitational wave observations of BH mergers provide independent probes of near-horizon geometry that could reveal rainbow modifications to the ringdown spectrum. The universal extremal deflection formula (Eq.~\eqref{deflection_extremal}) is particularly attractive for observational tests, as it eliminates the degeneracy between rainbow parameters that otherwise complicates interpretation.

Several directions merit further investigation building on this work. First, extending the analysis to rotating BHs described by Kerr-Newman-AdS metrics with rainbow modifications would capture the spin-dependent effects absent in the static configurations studied here. The interplay between frame-dragging and rainbow-induced modifications could produce distinctive polarization signatures in lensed electromagnetic radiation. Second, the strong deflection limit, where photons pass close to the PS, offers enhanced sensitivity to deviations from GR and warrants dedicated treatment using the logarithmic expansion pioneered by Bozza and collaborators. Third, incorporating thermodynamic considerations would establish connections between the lensing predictions and BH phase transitions in extended thermodynamic phase space, where the cosmological constant plays the role of pressure. Fourth, the quantum corrections encoded in rainbow functions could be connected more directly to specific UV-complete scenarios, such as loop quantum gravity or asymptotically safe gravity, providing theoretical grounding for the phenomenological parameters $f(\varepsilon)$ and $g(\varepsilon)$.

The reconciliation of WGC and WCCC demonstrated in this study relied on the rainbow-induced shift of the extremality bound. However, alternative mechanisms involving dark matter halos, quintessence fields, or modified gravity sectors could achieve similar effects through different physical pathways. Comparative studies across these scenarios would clarify which features are universal consequences of quantum gravity and which depend on specific model assumptions. The observational distinguishability of these mechanisms through their lensing signatures represents a promising target for theoretical development.

In summary, we established that gravity's rainbow provides a coherent framework for simultaneously satisfying WGC and WCCC constraints in charged AdS BHs. The PS analysis confirmed BH structure preservation across the parameter space, while the gravitational lensing formalism yielded explicit predictions for deflection angles in terms of rainbow functions. The universal $f$-independence of extremal lensing and the enhanced deflection from super-extremal configurations offer concrete observational targets. As precision gravitational-wave and electromagnetic observations probe ever closer to BH horizons, the signatures identified here position gravitational lensing as a powerful tool for testing quantum gravity phenomenology in astrophysical settings.

\section*{Acknowledgments}
\.{I}.~S. extends appreciation to T\"{U}B\.{I}TAK, ANKOS, and SCOAP3 for their financial assistance. Additionally, he acknowledges the support from COST Actions CA22113, CA21106, CA23130, and CA23115, which have been pivotal in enhancing networking efforts.

\end{document}